\input lanlmac.tex
\input epsf
\font\cmss=cmss10 \font\cmsss=cmss10 at 7pt
\ifx\epsfbox\UnDeFiNeD\message{(NO epsf.tex, FIGURES WILL BE IGNORED)}
\def\figin#1{\vskip2in}
\else\message{(FIGURES WILL BE INCLUDED)}\def\figin#1{#1}\fi
\def\tfig#1{{\xdef#1{Fig.\thinspace\the\figno}}
Fig.\thinspace\the\figno \global\advance\figno by1}

\def\subsubsection#1{

\vskip 0.3cm
{\it #1}
\vskip 0.2cm

}

\def\({\left(}
\def\){\right)}
\def\<{\left\langle\,}
\def\>{\, \right\rangle}


\def\Tr{\,{\rm Tr}\, }

\def\Im{\,{\rm Im}\, }
\def\Re{\,{\rm Re}\, }

\def\Det{{\rm Det }}

\def\({\left(}
\def\){\right)}
\def\[{\left[}
\def\]{\right]}
\def\p{\partial}

\def\11{1\!\! 1}

\def\hf{{1\over 2}}
\def\IR{\relax{\rm I\kern-.18em R}}
\def\a{\alpha}

\def\g{\gamma}

\def\th{\theta}

\def\u{\upsilon }

\def\X{\Xi}
\def\P{\Pi}

\def\P{\Phi}
\def\tP{\tilde\Phi}

\def\w{\omega}

 \def\IZ{\relax\ifmmode\mathchoice
{\hbox{\cmss Z\kern-.4em Z}}{\hbox{\cmss Z\kern-.4em Z}}
{\lower.9pt\hbox{\cmsss Z\kern-.4em Z}}
{\lower1.2pt\hbox{\cmsss Z\kern-.4em Z}}\else{\cmss Z\kern-.4em
Z}\fi}
\def\o{\omega }


\def\CB {{\cal B}}

\def\CF {{\cal F}}

\def\CO {{\cal O}}

\def\CZ {{\cal Z}}


\def\Sef{S_{\rm eff}}

 \def\R{\relax{\rm I\kern-.18em R}}
\font\cmss=cmss10 \font\cmsss=cmss10 at 7pt
\def\Z{\relax\ifmmode\mathchoice
{\hbox{\cmss Z\kern-.4em Z}}{\hbox{\cmss Z\kern-.4em Z}}
{\lower.9pt\hbox{\cmsss Z\kern-.4em Z}}
{\lower1.2pt\hbox{\cmsss Z\kern-.4em Z}}\else{\cmss Z\kern-.4em Z}\fi}

\def\np#1#2#3{{\it Nucl. Phys.} {\bf B#1} (#2) #3}
\def\pl#1#2#3{{\it Phys. Lett.} {\bf B#1} (#2) #3}
\def\prl#1#2#3{{\it Phys. Rev. Lett.} {\bf #1} (#2) #3}
\def\physrev#1#2#3{{\it Phys. Rev.} {\bf D#1} (#2) #3}

\def\jhep#1#2#3{{\it JHEP} {\bf #1} (#2) #3}
\def\hepth#1{{\tt hep-th}/#1}


\chardef\tempcat=\the\catcode`\@ \catcode`\@=11
\def\cyracc{\def\u##1{\if \i##1\accent"24 i%
    \else \accent"24 ##1\fi }}
\newfam\cyrfam
\font\tencyr=wncyr10
\def\cyr{\fam\cyrfam\tencyr\cyracc}
\def\frac#1#2{{\textstyle{#1\over#2}}}
\def\inv{^{\raise.15ex\hbox{${\scriptscriptstyle -}$}\kern-.05em 1}}


\lref\ZamolodchikovAH{
A.~B.~Zamolodchikov and A.~B.~Zamolodchikov,
``Liouville field theory on a pseudosphere,''
hep-th/0101152.
}

\lref\KazakovPM{
V.~Kazakov, I.~K.~Kostov and D.~Kutasov,
``A matrix model for the two-dimensional black hole,''
\np{622}{2002}{141},
hep-th/0101011.
}

\lref\ColemanAE{
S.~R.~Coleman,
``The Uses Of Instantons,''
HUTP-78/A004
{\it Lecture delivered at 1977 Int. School of Subnuclear Physics, Erice, Italy, Jul 23-Aug 10, 1977, Published in  Erice Subnucl.1977:0805}
}

\lref\EynardKF{
B.~Eynard,
JHEP {\bf 0311}, 018 (2003)
[arXiv:hep-th/0309036].
}

\lref\TeodorescuQM{
R.~Teodorescu, E.~Bettelheim, O.~Agam, A.~Zabrodin and P.~Wiegmann,
``Normal random matrix ensemble as a growth problem: 
Evolution of the spectral
curve,''
arXiv:hep-th/0401165.
}

\lref\SeibergShih{
N.~Seiberg and D.~Shih, ``Branes, rings and matrix models in minimal
(super)string theory,'' arXiv:hep-th/0312170.}

\lref\KostovHN{
I.~K.~Kostov,
``Strings embedded in Dynkin diagrams'',
In Cargese 1990, Proceedings, Random surfaces and quantum gravity, 135-149; ``Loop amplitudes for non-rational string theories'',
Phys.\ Lett.\ B266, 317 (1991).
}

\lref\MooreIR{
G.~W.~Moore, N.~Seiberg and M.~Staudacher,
``From loops to states in 2-D quantum gravity,''
Nucl.\ Phys.\ B {\bf 362}, 665 (1991).
}

\lref\AlexandrovNN{
S.~Y.~Alexandrov, V.~A.~Kazakov and D.~Kutasov,
``Non-perturbative effects in matrix models and D-branes,''
JHEP {\bf 0309}, 057 (2003)
[arXiv:hep-th/0306177].
}

\lref\DavidSK{
F.~David, ``Phases Of The Large N Matrix Model And
Nonperturbative Effects In 2-D Gravity,''
\np{348}{1991}{507}.
}

\lref\ChekhovGZ{
L.~Chekhov, A.~Marshakov, A.~Mironov and D.~Vasiliev,
``DV and WDVV,''
Phys.\ Lett.\ B {\bf 562}, 323 (2003)
[arXiv:hep-th/0301071].
}

\lref\DavidZA{
F.~David,
``Nonperturbative effects in matrix models and vacua
of two-dimensional gravity,''
\pl{302}{1991}{403}, hep-th/9212106.
}

\lref\KazakovYH{
V.~A.~Kazakov and A.~Marshakov,
``Complex curve of the two matrix model and its tau-function,''
J.\ Phys.\ A {\bf 36}, 3107 (2003)
[arXiv:hep-th/0211236]. }


%
\lref\DaulBG{
J.~M.~Daul, V.~A.~Kazakov and I.~K.~Kostov,
``Rational theories of 2-D gravity from the two matrix model,''
Nucl.\ Phys.\ B {\bf 409}, 311 (1993)
[arXiv:hep-th/9303093].  }

\lref\AKKnmm{
S.~Y.~Alexandrov, V.~A.~Kazakov and I.~K.~Kostov,
``2D string theory as normal matrix model,''
arXiv:hep-th/0302106. }

\lref\DouglasPT{
M.~R.~Douglas, ``The Two Matrix Model,'', in *Cargese 1990,
Proceedings, Random surfaces and quantum gravity* 77-83.}

\lref\PolchinskiFQ{
J.~Polchinski,
``Combinatorics Of Boundaries In String Theory,''
\physrev{50}{1994}{6041}, hep-th/9407031.
}
\lref\FUKUMA{
M. Fukuma, S. Yahikozawa,
``Comments on D-Instantons in $c<1$ Strings,
\pl{460}{1999}{71-78}, hep-th/9902169.}
\lref\NevesXT{
R.~Neves,
\pl{411}{1997}{73},
hep-th/9706069.
}


\lref\AlexandrovUN{
S.~Alexandrov,
``(m,n) ZZ branes and the c = 1 matrix model,''
arXiv:hep-th/0310135.
}


\lref\McGreevyKB{
J.~McGreevy and H.~Verlinde,
``Strings from tachyons: The c = 1 matrix reloated,''
hep-th/0304224.
}

\lref\CallanPT{
C.~G.~.~Callan and S.~R.~Coleman,
``The Fate Of The False Vacuum. 2. First Quantum Corrections,''
Phys.\ Rev.\ D {\bf 16}, 1762 (1977).
}

\lref\MartinecKA{
E.~J.~Martinec,
``The annular report on non-critical string theory,''
arXiv:hep-th/0305148.
}

\lref\KlebanovKM{
I.~R.~Klebanov, J.~Maldacena and N.~Seiberg,
``D-brane decay in two-dimensional string theory,''
hep-th/0305159.
}

\lref\AlexandrovUT{
S.~Alexandrov,
 ``Matrix quantum mechanics and two-dimensional string theory in non-trivial
backgrounds,''
arXiv:hep-th/0311273.
}

\lref\TeschnerQK{
J.~Teschner,
arXiv:hep-th/0308140.
}

\lref\PZinnB{P. Zinn-Justin,
``Universality of correlation functions of hermitian random 
matrices in an   external field", Commun. Math. Phys. 194, 631-650 (1998), 
 cond-mat/9705044.}

\lref\ZinnJustinEM
\lref\ZinnJustinEM{
P.~Zinn-Justin,
``Random hermitian matrices in an external field,''
Nucl.\ Phys.\ B {\bf 497}, 725 (1997)
[arXiv:cond-mat/9703033].
}

\lref\KazakovAZ{
V.~A.~Kazakov and T.~Wynter,
``Large N phase transition in the heat kernel on the U (N) group,''
Nucl.\ Phys.\ B {\bf 440}, 407 (1995)
[arXiv:hep-th/9410087].
}

\lref\KazakovAE{
V.~A.~Kazakov, M.~Staudacher and T.~Wynter,
``Character expansion methods for matrix models of dually weighted graphs,''
Commun.\ Math.\ Phys.\  {\bf 177}, 451 (1996)
[arXiv:hep-th/9502132].
}

\lref\KazakovGM{
V.~A.~Kazakov, M.~Staudacher and T.~Wynter,
``Almost flat planar diagrams,''
Commun.\ Math.\ Phys.\  {\bf 179}, 235 (1996)
[arXiv:hep-th/9506174].
}

\lref\KazakovZM{
V.~A.~Kazakov, M.~Staudacher and T.~Wynter,
``Exact Solution of Discrete Two-Dimensional R~2 Gravity,''
Nucl.\ Phys.\ B {\bf 471}, 309 (1996)
[arXiv:hep-th/9601069].
}

\lref\KlebanovWG{
I.~R.~Klebanov, J.~Maldacena and N.~Seiberg,
``Unitary and complex matrix models as 1-d type 0 strings,''
arXiv:hep-th/0309168.
}

\lref\DouglasUP{
M.~R.~Douglas, I.~R.~Klebanov, D.~Kutasov, J.~Maldacena, E.~Martinec and N.~Seiberg,
``A new hat for the c = 1 matrix model,''
arXiv:hep-th/0307195.
}

\lref\EynardSG{
B.~Eynard and J.~Zinn-Justin,
``Large order behavior of 2-D gravity coupled to $d < 1$ matter,''
\pl{302}{1993}{396}, hep-th/9301004.
}

\lref\DijkgraafHK{
R.~Dijkgraaf, G.~W.~Moore and R.~Plesser,
``The Partition function of 2-D string theory,''
\np{394}{1993}{356},
hep-th/9208031.
}

\lref\GinspargCY{
P.~Ginsparg and J.~Zinn-Justin,
``Action Principle And Large Order Behavior Of Nonperturbative Gravity,''
LA-UR-90-3687
{\it Lectures to appear in Proc. of Cargese Workshop:
Random Surfaces and Quantum Gravity, Ed. by O. Alvarez, et al., Cargese,
France, May 27 - June 2, 1990}
}

\lref\ItzyksonFI{
C.~Itzykson and J.~B.~Zuber,
``The Planar Approximation. 2,''
J.\ Math.\ Phys.\  {\bf 21}, 411 (1980).
}

\lref\CardyIR{
J.~L.~Cardy,
``Boundary Conditions, Fusion Rules And The Verlinde Formula,''
Nucl.\ Phys.\ B {\bf 324}, 581 (1989).
}

\lref\ZamolodchikovAA{
A.~B.~Zamolodchikov and A.~B.~Zamolodchikov,
``Structure constants and conformal bootstrap in Liouville field theory,''
\np{477}{1996}{577},
hep-th/9506136.
}
\lref\DornXN{
H.~Dorn and H.~J.~Otto,
``Two and three point functions in Liouville theory,''
\np{429}{1994}{375},
hep-th/9403141.
}

\lref\DouglasVE{
M.~R.~Douglas and S.~H.~Shenker,
``Strings In Less Than One-Dimension,''
\np{335}{1990}{635}.
}

\lref\DouglasDD{
M.~R.~Douglas,
``Strings In Less Than One-Dimension And The Generalized K-D-V Hierarchies,''
Phys.\ Lett.\ B {\bf 238}, 176 (1990).
}

\lref\BrezinRB{
E.~Brezin and V.~A.~Kazakov,
``Exactly Solvable Field Theories Of Closed Strings,''
\pl{236}{1990}{144}.
}
\lref\GrossVS{
D.~J.~Gross and A.~A.~Migdal,
``Non-perturbative Two-Dimensional Quantum Gravity,''
\prl{64}{1990}{127}.
}

\lref\WittenZD{
E.~Witten,
``Ground ring of two-dimensional string theory,''
Nucl.\ Phys.\ B {\bf 373}, 187 (1992)
[arXiv:hep-th/9108004].
}

\lref\KricheverCI{
I.~Krichever, M.~Mineev-Weinstein, P.~Wiegmann and A.~Zabrodin,
``Laplacian Growth and Whitham Equations of Soliton Theory,''
arXiv:nlin.si/0311005.
}

\lref\KazakovHY{
V.~A.~Kazakov,
``Exact Solution Of The Ising Model On A Random Two-Dimensional Lattice,''
JETP Lett.\  {\bf 44}, 133 (1986)
[Pisma Zh.\ Eksp.\ Teor.\ Fiz.\  {\bf 44}, 105 (1986)].
}

\lref\BoulatovSB{
D.~V.~Boulatov and V.~A.~Kazakov,
 ``The Ising Model On Random Planar Lattice: The Structure Of Phase Transition
And The Exact Critical Exponents,''
Phys.\ Lett.\  {\bf 186B}, 379 (1987).
}

\lref\StaudacherXY{
M.~Staudacher,
``Combinatorial solution of the two matrix model,''
Phys.\ Lett.\ B {\bf 305}, 332 (1993)
[arXiv:hep-th/9301038].
}

\lref\EynardKG{
B.~Eynard,
``Large N expansion of the 2-matrix model,''
JHEP {\bf 0301}, 051 (2003)
[arXiv:hep-th/0210047].
}

\lref\DijkgraafFC{
R.~Dijkgraaf and C.~Vafa,
``Matrix models, topological strings, and supersymmetric gauge theories,''
Nucl.\ Phys.\ B {\bf 644}, 3 (2002)
[arXiv:hep-th/0206255].
}

\lref\FZZ{V. Fateev, A. Zamolodchikov and  Al. Zamolodchikov,
unpublished.}

\lref\AlexandrovUN{
S.~Alexandrov,
``$(m,n)$ ZZ branes and the c = 1 matrix model,''
arXiv:hep-th/0310135.
}

\lref\Dijkgraafcar{R. Dijkgraaf, ``Intersection Theory, Integrable Hierarchies and Topological Field Theory", 
 Lectures given at the   Cargese Summer School on `New Symmetry Principles in Quantum Field Theory,'   July 16-27, 1991, 
 hep-th/9201003.}

\lref\MatytsinIQ{
A.~Matytsin,
``On the large N limit of the Itzykson-Zuber integral,''
Nucl.\ Phys.\ B {\bf 411}, 805 (1994)
[arXiv:hep-th/9306077].
}

\lref\Hos{
K.~Hosomichi,
``Bulk-Boundary Propagator in Liouville Theory on a Disk,''
\jhep{0111}{2001}{044},
hep-th/0108093.
}

\lref\Pons{
B.~Ponsot,
``Liouville Theory on the Pseudo-sphere: Bulk-Boundary Structure Constant,''
hep-th/0309211.
}

\lref\TeschnerQK{
J.~Teschner,
 ``On boundary perturbations in Liouville theory and brane dynamics in
noncritical string theories,''
arXiv:hep-th/0308140.
}

\lref\DiFrancescoJD{
P.~Di Francesco and D.~Kutasov,
``Unitary Minimal Models Coupled To 2-D Quantum Gravity,''
Nucl.\ Phys.\ B {\bf 342}, 589 (1990).
}

\lref\difkut{P. Di Francesco and D. Kutasov, hep-th/9109005,
Nucl. Phys. {\bf B375} (1992) 119.}
\lref\FATA{ L. D. Faddeev and L. A. Takhtajan, ``Hamiltonian Methods in
the Theory of Solitons'', Springer-Ferlag (1987).}
\lref\SHENKER{S. Shenker, Proceedings of Cargese workshop
on Random Surfaces, Quantum Gravity and Strings, 1990.}

\lref\BertolaKZ{
M.~Bertola,
``Second and third order observables of the two-matrix model,''
JHEP {\bf 0311}, 062 (2003)
[arXiv:hep-th/0309192].
}

\lref\RecknagelRI{
A.~Recknagel, D.~Roggenkamp and V.~Schomerus,
``On relevant boundary perturbations of unitary minimal models,''
Nucl.\ Phys.\ B {\bf 588}, 552 (2000)
[arXiv:hep-th/0003110].
}

\lref\BehrendBN{
R.~E.~Behrend, P.~A.~Pearce, V.~B.~Petkova and J.~B.~Zuber,
``Boundary conditions in rational conformal field theories,''
Nucl.\ Phys.\ B {\bf 570}, 525 (2000)
[Nucl.\ Phys.\ B {\bf 579}, 707 (2000)]
[arXiv:hep-th/9908036].
}


 \lref\AlexandrovNN{ S.~Y.~Alexandrov, V.~A.~Kazakov and
D.~Kutasov, ``Non-Perturbative Effects in Matrix Models and
D-branes,''  hep-th/0306177.
}

\lref\DiFrancescoGinsparg{P.~Di Francesco, P.~Ginsparg and
J.~Zinn-Justin,``2-D Gravity and random matrices,'' Phys.\ Rept.\
{\bf 254} (1995) 1,  hep-th/9306153.
}

\lref\polchinski{J. Polchinski, ``What is string theory'',
{\it Lectures presented at the 1994 Les Houches Summer School
``Fluctuating Geometries in Statistical Mechanics and Field Theory''},  
\hepth{9411028}.}

\lref\KlebanovMQM{I. Klebanov, {\it Lectures delivered at the ICTP
Spring School on String Theory and Quantum Gravity},
Trieste, April 1991, \hepth{9108019}.}

\lref\SchomerusVV{ V.~Schomerus, ``Rolling tachyons from Liouville
theory,'' arXiv:hep-th/0306026.
}

\lref\Icar{
I. Kostov , ``Solvable Statistical Models on Random
Lattices",Proceedings
 of the Conference on recent developments in statisticalmechanics and
quantum field theory.
  (Trieste, 10 - 12 April 1995),Nucl. Phys. B (Proc. Suppl.)
 45 A (1996) 13-28, hep-th/9509124 .
}

\lref\McGreevyEP{
J.~McGreevy, J.~Teschner and H.~Verlinde,
``Classical and quantum D-branes in 2D string theory,''
hep-th/0305194.
}

\lref\TeschnerQK{ J.~Teschner, ``On boundary perturbations in
Liouville theory and brane dynamics in noncritical string
theories,''  hep-th/0308140.
}

\lref\DV{R. Dijkgraaf,  C. Vafa, 
``N=1 Supersymmetry, Deconstruction, and Bosonic Gauge Theories'',
hep-th/0302011.
}
 
  \lref\kkk{V. Kazakov,  I. Kostov,  D.  Kutasov,
  ``A Matrix Model for the Two Dimensional Black Hole",
  Nucl.Phys. B622 (2002) 141,
  hep-th/0101011.
  }

   \lref\JevickiQN{
A.~Jevicki,
``Developments in 2-d string theory,''
 hep-th/9309115.
}

\lref\Sennew{ A.~Sen, ``Open-Closed Duality: Lessons from the
Matrix Model,''  hep-th/0308068.
}
   \lref\Kstau{I. Kostov and 
   M. Staudacher, ``Strings in discrete and continuous target 
 spaces:a comparison", \pl{305}{1993}{43}.
   }
   \lref\FZZb{V.~Fateev, A.~B.~Zamolodchikov and A.~B.~Zamolodchikov,
``Boundary Liouville field theory. I: Boundary state and boundary
two-point function,''  hep-th/0001012.
}

 \lref\DO{
H. Dorn, H.J. Otto: 
Two and three point functions in Liouville theory, 
\np{429}{1994}{375-388}.
}
 
\lref\PTtwo{B.~Ponsot, J.~Teschner, ``Boundary Liouville Field 
Theory: Boundary three point function'',
  Nucl.~Phys.~{B622} (2002) 309, \hepth{0110244}.
  }

\lref\Teschner{
J.Teschner. On the Liouville Three-Point Function.
Phys.Lett., B363 (1995) 65. 
}
  \lref\ReSch{A. Recknagel,  V. Schomerus,
  ``Boundary Deformation Theory and Moduli Spaces of D-Branes",
   Nucl.Phys. B545 (1999) 233, hep-th/9811237.
  }  

  \lref\DiK{
  P. Di Francesco,  D. Kutasov, 
  ``World Sheet and Space Time Physics in Two Dimensional (Super)
 String   Theory", Nucl.Phys. B375 (1992) 119,   hep-th/9109005.
}
  \lref\KMSnew{
I. R. Klebanov,  J. Maldacena,  N. Seiberg,
``Unitary and Complex Matrix Models as 1-d Type 0 Strings'',
   hep-th/0309168.
   }
  \lref\GM{
  P. Ginsparg and G. Moore,
  ``Lectures on 2D gravity and 2D string theory (TASI 1992)", 
  hep-th/9304011.
   }

\lref\Gouli{
M. Goulian and B. Li, Phys. Rev. Lett. 66 (1991), 2051.
}
 
\lref\VDotsenko{
V. Dotsenko, Mod. Phys. Lett. A6(1991), 3601.
}

  \lref\Witten{
  E.~Witten,
``Ground ring of two-dimensional string theory,''
Nucl.\ Phys.\ B {\bf 373}, 187 (1992),
hep-th/9108004.
}

\lref\WitZw{ E.~Witten and B.~Zwiebach, ``Algebraic structures
and differential geometry in 2D string theory,'' Nucl.\ Phys.\
B {\bf 377}, 55 (1992),  hep-th/9201056.
}

 \lref\KMS{ David Kutasov, Emil J. Martinec, Nathan Seiberg, 
   ``Ground rings and their modules in 2-D gravity with $c\le1 $ matter",  
      Phys.Lett. {\CB276} (1992) 437, \hepth{9111048}.
}

 \lref\KlebanovMQM{
  I.~R.~Klebanov, ``String theory in two-dimensions,''
 hep-th/9108019.
}
  \lref\bershkut{
  M. Bershadsky and D. Kutasov,
 ``Scattering of open and closed strings in (1+1)-dimensions",
 \np{382}{1992}{213}, \hepth{9204049}.
  }
  \lref\berkut{
  M. Bershadsky and D. Kutasov,
  Phys. Lett. 274B (1992) 331.
  }

 \lref\Idis{I. Kostov, ``Strings with discrete target space'',
 \np{376}{1992}{539},  hep-th/9112059.
}

\lref\KKloop{
V. Kazakov, I. Kostov,
``Loop Gas Model for Open Strings'',
 \np{386}{1992}{520}.
}
 \lref\XID{Xi Yin, 
 ``Matrix Models, Integrable Structures, and T-duality of Type 0 String   Theory",
 hep-th/0312236.
 }  
\lref\Ibliou{
I. K. Kostov, ``Boundary Correlators in 2D Quantum 
Gravity:
 Liouville versus Discrete Approach ", \np{658}{2003}{397},
 hep-th/0212194.}
 \lref\KPS{I. Kostov, B. Ponsot and D. Serban,   ``Boundary  Liouville Theory and 
2D Quantum Gravity", hep-th/0307189.}

\lref\ZZPseudo{
A.~B. Zamolodchikov and A.~B. Zamolodchikov, ``Liouville field theory on a
  pseudo-sphere,  
  \hepth{0101152}.
}

\lref\GovJcf{
S. Govindarajan, T. Jayaraman and V. John,
ÒGenus Zero Correlation Functions in $c<1$ String Theory,Ó Phys. Rev. D 48 (1993) 839,
\hepth{9208064}.
}

\lref\MartinecKA{ E.~J.~Martinec, ``The annular report on non-critical string
theory,''  hep-th/0305148.}

  \lref\GovLast{
  S. Govindarajan, T. Jayaraman and V. John,
  ``Correlation Functions and Multi-critical Flows in $c<1$ String Theory",
 Int. J. Mod.Phys. A10 (1995) 477,  hep-th/9309040.}
     
   \lref\TY{Y. Tanii, S.-I. Yamaguchi, ``Two-dimensional 
   quantum gravity on a disc", Mod.Phys.Lett.A7 (1992) 521,
 hep-th/9110068; ``Disk Amplitudes in Two-Dimensional Open String Theories", hep-th/9203002.
  }
 \lref\ValyaJB{
 V.B. Petkova,  J.-B. Zuber, ``BCFT: from the boundary to the bulk'',
 Talk presented at TMR-conference "Non-perturbative Quantum Effects   2000", 
hep-th/0009219.
  }
 \lref\higkos{
 S. Higuchi and  I. Kostov , ``Feynman rules for   
string 
theories with 
discrete target space", \pl{357}{1995}{62}, hep-th/9506022. }
 \lref\topint{M. Aganagic,  R. Dijkgraaf,  A. Klemm,  M. Marino,  C. Vafa,
 ``Topological Strings and Integrable Hierarchies'',
 hep-th/0312085.
 }

\lref\newhat{
M.~R.~Douglas, I.~R.~Klebanov, D.~Kutasov, J.~Maldacena, E.~Martinec and N.~Seiberg,
``A new hat for the c = 1 matrix model,''
arXiv:hep-th/0307195.}

   \lref\KlebPol{
I.~R.~Klebanov and A.~M.~Polyakov, ``Interaction of discrete
states in two-dimensional string theory,'' Mod.\ Phys.\ Lett.\ A
{\bf 6}, 3273 (1991) hep-th/9109032.
}

 \lref\insts{S.  Alexandrov,  "D-branes and complex curves in $c=1$ string theory",
 hep-th/0403116.}

   \lref\DMP{ 
R.~Dijkgraaf, G.~W.~Moore and R.~Plesser,
``The Partition function of 2-D string theory,''
Nucl. Phys. B {394}, 356 (1993)
 hep-th/9208031.}
 
\lref\AKK{
S.  Alexandrov,  V. Kazakov,  I.  Kostov, 
``Time-dependent backgrounds of 2D string theory",
  Nucl.Phys. B640 (2002) 119,  hep-th/0205079.
  }
  \lref\Iflows{I. Kostov, ``Integrable flows in $c=1$ string theory",
  J.Phys. A36 (2003) 3153,  hep-th/0208034.}
  
  \lref\adem{ I. Kostov, 
  ``Gauge Invariant Matrix Model for the \^A-\^D-\^E Closed Strings",
  Phys.Lett. B297 (1992) 74-81, hep-th/9208053.
  }
\lref\KostovIE{
I.~K.~Kostov,
``Gauge invariant matrix model for the \^A-\^D-\^E  closed strings,''
Phys.\ Lett.\ B {\bf 297}, 74 (1992)
[arXiv:hep-th/9208053].
}

   \lref\ADKMV{
  M.~Aganagic, R.~Dijkgraaf, A.~Klemm, M.~Marino and C.~Vafa,
``Topological Strings and Integrable Hierarchies,''  hep-th/0312085.
  }
 \lref\SeibergS{
 N. Seiberg and D. Shih, ``Branes, rings and matrix models on minimal (super)string theory",  hep-th/0312170.
 }

\lref\Iham{
I. Kostov,
``Loop space Hamiltonian for $c \le 1$ open strings'',
Phys.Lett. B349 (1995) 284, hep-th/9501135.
}

\lref\Iopen{
I. Kostov,
``Field Theory of Open and Closed Strings with Discrete Target Space'',
Phys.Lett. B344 (1995) 135,
    hep-th/9410164.}

\lref\McGreevyKB{
J.~McGreevy and H.~Verlinde,
``Strings from tachyons: The c = 1 matrix reloaded,''
JHEP {\bf 0312}, 054 (2003)
[arXiv:hep-th/0304224].
}

\lref\McGreevyEP{
J.~McGreevy, J.~Teschner and H.~Verlinde,
``Classical and quantum D-branes in 2D string theory,''
JHEP {\bf 0401}, 039 (2004)
[arXiv:hep-th/0305194].
}

\lref\KricheverQE{
I.~M.~Krichever,
 ``The tau function of the universal Whitham hierarchy, matrix models and
topological field theories,''
arXiv:hep-th/9205110.
}

\lref\KlebanovKM{
I.~R.~Klebanov, J.~Maldacena and N.~Seiberg,
``D-brane decay in two-dimensional string theory,''
JHEP {\bf 0307}, 045 (2003)
[arXiv:hep-th/0305159].
}

\lref\KharchevKD{
S.~Kharchev and A.~Marshakov, ``Topological versus nontopological
theories and p - q duality in $c\le 1$ 2-d gravity models,''
arXiv:hep-th/9210072.
}

\lref\KharchevAS{
S.~Kharchev and A.~Marshakov,
``On p - q duality and explicit solutions in $c\le 1$ 2-d gravity models,''
Int.\ J.\ Mod.\ Phys.\ A {\bf 10}, 1219 (1995)
[arXiv:hep-th/9303100].
}

\lref\MarshakovNZ{
A.~Marshakov,
``String theory and classical integrable systems,''
arXiv:hep-th/9404126.
}

\lref\KharchevCP{
S.~Kharchev, A.~Marshakov, A.~Mironov and A.~Morozov,
``Landau-Ginzburg topological theories in the framework of GKM and equivalent
hierarchies,''
Mod.\ Phys.\ Lett.\ A {\bf 8}, 1047 (1993)
[Theor.\ Math.\ Phys.\  {\bf 95}, 571 (1993\ TMFZA,95,280-292.1993)]
[arXiv:hep-th/9208046].
}

\overfullrule=0pt
\Title{{hep-th/0403152}\hfill\vbox{\baselineskip12pt\hbox
{SPhT-04/026}\hbox{LPTENS-04/10 }}
}
{\vbox{\centerline
 { Instantons in  Non-Critical Strings   }
 \centerline{ from the  Two-Matrix Model }
\centerline{ }
 \vskip-2pt
}}
 %
 
  \vskip-20pt
 \centerline{
Vladimir A. Kazakov
$^{1}$\footnote{$^\bullet$}{Membre de l'Institut Universitaire de France}
 and Ivan K. Kostov$^2$\footnote{$^{\circ}$}{{ 
Associate member of the  Institute for Nuclear Research and Nuclear Energy
  {\cyr (IYaIYaE)},  72 Tsarigradsko Chaussee,
1784 Sofia, Bulgaria }}
 }

{ \ninepoint
\centerline{$^1${\it  Laboratoire de Physique Th\'eorique de l'Ecole
Normale Sup\'erieure\footnote{$^\ast$}{Unit\'e mixte de Recherche du
Centre National de la Recherche Scientifique et de  l'Ecole Normale
Sup\'erieure associ\'ee \`a l'Universit\'e de Paris-Sud et 
l'Universit\'e  Paris-VI
}  et
}}
}
\vskip -6pt
\centerline{{\it l'Universit\'e  Paris-VI, 24 rue Lhomond, 75231 Paris Cedex, France}}

\centerline{$^2${\it  Service de Physique 
Th{\'e}orique, CNRS -- URA 2306, }} 
\vskip -6pt
\centerline{{\it C.E.A. - Saclay,
F-91191 Gif-Sur-Yvette  Cedex, France}}

 
\vskip0.7cm

We derive the non-perturbative corrections to the free energy of the
two-matrix model in terms of its algebraic curve.  The eigenvalue
instantons are associated with the vanishing cycles of the curve. For
the $(p,q)$ critical points our results agree with the geometrical
interpretation of the instanton effects recently discovered in the CFT
approach.  The form of the instanton corrections implies that the
linear relation between the FZZT and ZZ disc amplitudes is a general
property of the 2D string theory and holds for any classical
background.  We find that the agreement with the CFT results holds in
presence of infinitesimal perturbations by order operators and observe
that the ambiguity in the interpretation of the eigenvalue instantons
as ZZ-branes (four different choices for the matter and Liouville
boundary conditions lead to the same result) is not lifted by the
perturbations.  We find similar results to the $c=1$ string
theory in
presence of tachyon perturbations.

\Date{March, 2004
\hfill {\it Dedicated to the Memory of Ian Kogan }}

%

\baselineskip=14pt plus 2pt minus 2pt

\newsec{Introduction } %

 \noindent 
  The non-perturbative  phenomena  observed in the early 90's in the solvable 
  models of non-critical string theories \refs{\BrezinRB\DouglasVE\DouglasDD\GrossVS\DavidSK\DavidZA-\GinspargCY}  and studied further in \refs{\EynardSG, \FUKUMA}
are now much better understood 
 thanks to the recent regain of interest in this subject.
  In the papers \refs{\McGreevyKB\MartinecKA\KlebanovKM\DouglasUP-\AlexandrovNN},
 the non-perturbative corrections to the string partition function 
 were given a world sheet interpretation in terms of amplitudes of
 open strings attached to ZZ branes discovered by A. and
 Al. Zamolodchikov\ZamolodchikovAH.
 The  agreement between the  matrix  and  CFT descriptions 
 of the non-perturbative phenomena was observed  
 not only the minimal $(p,q)$ models of 2D quantum
 gravity but also the $c=1$ string theory in the presence of vortex
 perturbations \refs{\AlexandrovNN,\AlexandrovUN},  which is believed 
 to describe the 2D
 black hole \KazakovPM.

As it was already pointed out in
\ZamolodchikovAH\ and then elaborated in 
\refs{\MartinecKA, \McGreevyEP, \TeschnerQK, \KlebanovWG}, the
 disc amplitudes on ZZ branes appeared to be the same as the disc
 amplitudes on FZZT branes taken at special complex values of the
 boundary cosmological coupling. 
A  deeper geometrical  understanding  of this relation 
from the CFT side was   
achieved in   the recent
work of Seiberg and Shih \SeibergShih\ where the CFT description of the $(p,q)$ 
models  based on the ground ring structure
\refs{\Witten, \KMS}.   It was shown that the ground ring  relations 
lead to the same algebraic curve 
that appears in the matrix models  approach.
  The algebraic curve found in \SeibergShih\  gives  the same
   representation of the  correlation functions of the  the FZZT brane 
   in terms of Chebyshev   polynomials
 as the one found   in the   loop gas models in \KostovHN\   and later 
 in the  two-matrix model in  \refs{\EynardSG, \DaulBG}. 
   The authors of   \SeibergShih\     
  gave a nice interpretation of the ZZ branes 
   as degenerate (pinched)   cycles   of the complex curve.

 The disc  amplitude   on   the ZZ brane  associated with  the degenerate cycle 
 $A_{mn}$ is  given  by  the contour integral  of  a  certain holomorphic
differential   along    the   dual  cycle $B_{mn}$.
The $c=1$  version of this geometrical  interpretation  was proposed in 
\insts.

 In the matrix approach the non-perturbative effects are 
produced by eigenvalue tunneling amplitudes. Since the 
complex curve   is determined by  the shape of the effective potential, it
is natural to seek a correspondence between the ZZ branes and the
 ``non-minimal" saddle points  associated with the 
  local extrema  of the effective potential.

In this paper we   show  that   the ``non-minimal" saddle points 
 are associated with the pinched cycles of the complex curve and 
  express the instanton amplitudes through integrals  along
  the dual cycles.  
   Our results show that the   geometrical picture found in    \SeibergShih\ 
 for the  $(p,q)$ critical points, actually  holds for a  general string background. 
 
  We will consider the example of the  two matrix model
(2MM).  It is known that  all   rational $(p,q)$ models  can be obtained by an appropriate
tuning of the potentials in 2MM \refs{\DouglasPT,\DaulBG}.   
The 2MM gives probably the most economical matrix description of all
non-critical string theories with $c\le 1$ matter content, such as
pure gravity (c=0), exact solution of Ising on random dynamical graphs
(c=1/2) \refs{\KazakovHY,\BoulatovSB} or the $2D$ string theory on the
self-dual radius \refs{\DMP\MI-\AKKnmm}.  In case of a a polynomial potential, the planar limit  of this model   is described  in terms of an algebraic (in general non-hyper-elliptic)  curve \refs{\StaudacherXY\EynardKG-\KazakovYH}.

 We  evaluate the non-perturbative effects in the 2MM  at the $(p,p+1)$  
 critical point  by  extending the  quasi-classical   analysis  developed  by  F. David 
  \DavidZA\  for   the 1MM.    
    In order to obtain  dimensionless 
  quantities, we also  calculate the normalized  free energy   for given 
  complex curve  
  using the Riemann bilinear identity.
   The results reproduce those  obtained by  Eynard and Zinn-Justin
   \EynardSG\  from the  string equation   in the double scaling limit.

 
 We also consider the small perturbations of the algebraic curve 
 around the $(p,p+1)$ critical point generated by order operators.  We
 evaluate from the 2MM the leading order non-perturbative effects in
 the perturbed theory and compare them with the predictions of the
 boundary CFT which we extract from the known one point functions on
 the ZZ brane (normalized by the two point functions on the sphere to
 compare the dimensionless ratios). In the non-perturbed theory, there
 are four possible choices for the Liouville and matter boundary
 conditions, which lead to the same result.  We found that this
 degeneracy is not lifted by perturbations by order operators, at
 least in linear order in the couplings.  All four choices give to the
 same expression, which agrees with the one obtained in 2MM.
  
 Finally we discuss the $c\to 1$ limit and the instanton corrections
 in the Matrix Quantum Mechanics. We show that the instanton
 amplitudes in presence of vortex condensation, obtained previously
 using the equations of Toda hierarchy \refs{\AlexandrovNN,
 \AlexandrovUN} can be obtained as integrals along compact cycles of
 the complex curve, very much as in the case of the $c<1$ theory.

\newsec{ Spectral curve, free energy and instantons
of the general 2MM }

 \noindent 
In this section we will review the geometrical description of the
planar limit of the 2MM \refs{\EynardKG,\KazakovYH}
 including the
formulas for the free energy in terms of the integrals of a  certain
holomorphic differential along the closed A and B cycles on the
algebraic curve of the model.  We also will give  expressions for the
 non-perturbative corrections 
 in terms of similar integrals  along  
non-trivial cycles   passing through  conical singularities, or  double points, 
of the  the algebraic curve\foot{If the curve is defined by the equation 
 $F(x,y)=0$, then the  double points are those for which 
 $dF(x,y)=0$.}.
   The  conical singularities  can   
   be  interpreted,  following     \SeibergShih, as   
  degenerate, or ``pinched", cycles  of a curve  of higher genus.   
These results will be used 
   in   the next section, where the $(p,q)$ critical regimes of the
2MM will be considered.

\subsec{Effective potential and   saddle-point equations}

 \def\X{ {\bf X}}
 \def\Y{ {\bf Y}}

  \noindent 
 The partition function of the 2MM is defined as  
 \eqn\partF{
\CZ_N=\int d\X\ d\Y \ e^{   \Tr [ \X\Y - V(\X)- \tilde V(\Y)] }
 \equiv \<1\>_N }
 where   $\X,\Y$
  are the $N\times N$ hermitian matrices
 and
\eqn\POTAB{ \eqalign{ 
V(\X)&=\sum_{k=1}^{q} T_k \X^k ,\qquad 
\tilde V(\Y)=\sum_{k=1}^{p} \tilde T_k \Y^k  }}
are polynomial potentials.  
Using the representation of the 2MM in terms of the eigenvalues
$x_k,y_k$ \ItzyksonFI
\eqn\EIGR{
\CZ_N=\int \prod_{k=1}^N\( dx_k \ dy_k \ 
e^{   x_ky_k - V(x_k)- \tilde V(y_k)}\) \Delta_N(x)\Delta_N(y)
 }
where $\Delta_N(z)=\prod_{k>j=1}^N (z_k-z_j)$ is the Vandermonde
determinant, the ratio of partition functions $\CZ_{N+1}$ and
$\CZ_{N}$ can be expressed as the double   integral
\eqn\CHAN{  
{\CZ_{N+1}\over \CZ_N}=  \int\limits_{-\infty}^\infty \int\limits_{-\infty}^\infty
dxdy\ e^{-    S_{\rm eff}(x,y)}  ,
}
where 
$\Sef(x,y)$ is  the effective action  for a pair of eigenvalues $x,y$:
\eqn\seff{ 
\Sef(x,y)=- xy  +V(x)+\tilde V(y) -\log  \< \Det (x-\X) \, \Det (y-\Y)    \>_N.
}
The integrand in the r.h.s. of \CHAN \
  is the expectation value of having
one eigenvalue of the matrix $\X$ at position $x$ and one eigenvalue of
the matrix $\Y$ at position $y$.
In the large $N$ limit one can use
 the factorization properties $ \< \Det  A  \, \Det  B \>= 
 \< \Det  A \>\< \Det  B \>$ and 
$ \log\< \Det  A\>=\< \Tr\log A\>$, 
 to   write the effective action in the form 
 \eqn\seff{ 
\Sef(x,y)=- xy+  \P(x)+\tP(y), 
}
 with
\eqn\AVLOG{\eqalign{ \P(x)&= V(x)- \<\Tr\log (x-\X) \>_N \cr
 \tP(y)&= \tilde
V(y)- \<\Tr\log (y-\Y) \>_N}
}
and calculate the double integral \CHAN\ by the saddle point method.
 The  saddle point equations
\eqn\Sdpt{ \eqalign{x= \Phi '(y),\qquad  y=\tilde \Phi'(x) }
}
  determine  not only the position of the saddle points,  but also   the 
  potentials \AVLOG\ themselves.    
As  follows from the   studies   of the two-matrix model 
 by different techniques 
\refs{ \DaulBG,   \EynardKG, \MatytsinIQ,\ZinnJustinEM }
\foot{ A  similar approach to the evaluation of the large $N$ characters and heat kernels  is used in 
\refs{\KazakovAZ\KazakovAE\KazakovGM-\KazakovZM}.}, the
solution of the model in the planar limit can be expressed in terms of
a functional dependence between the complex variables $x$ and $y$:
\eqn\XYxy{ \eqalign{x= X(y),\qquad  y=Y(x)}
}
where the functions $X$ and $Y$ are  inverse to each other if 
considered as {multivalued} meromorphic functions defined on their Riemann surfaces.  On the physical sheets   $Y(x)=\Phi'(x)$ and $X(y) = 
\tilde\Phi'(y)$ and therefore they
  satisfy  (again on the  physical sheets)  the
asymptotic relations

\eqn\ASYM{\eqalign{
Y(x)&=V'(x)-  N/x + o(1/x^2), \qquad x\to\infty \cr
X(y)&=\tilde V'(y)-N/y + o(1/y^2), \qquad  y\to\infty.
} }
These conditions and the fact that the two meromorphic 
functions \XYxy\ are inverse to each other 
and have no other  poles except those  at infinity  determine them completely.

%
        
%


\subsec{Spectral curve}

  The geometrical object behind is a complex curve $F(x,y)=0$ whose
  projections to the $x$ and $y$ complex planes are given,
  correspondingly, by the Riemann surfaces of the meromorphic
  functions $y=Y(x)$ and $x=X(y)$. 
  For polynomial potentials \POTAB,
  the curve is algebraic and its equation 
  \refs{\StaudacherXY\EynardKG-\KazakovYH} is
  a direct consequence of
  the asymptotics \ASYM  :

\eqn\CURVE{ F(x,y)\equiv \[y-V'(x)\][x-\tilde
V'(y)]+P (x,y)=0 ,}
 where $P (x,y)$ is a polynomial  of degree $(q-2, p-2)$   \EynardKG
\eqn\PTR{ P (x,y)= \<\Tr {V'(x)-V'(\X)\over x-\X} {\tilde
V'(y)-\tilde V'(\Y)\over y-\Y}\>-N. }
The coefficients
of the polynomial $P(x,y)$ 
 (the moduli   of the spectral curve)
are determined by the potentials \POTAB\ 
 through the asymptotic conditions \ASYM\  as well as by the
filling numbers  $N_1, N_2,...$ $(\sum_{i} N_i=N)$
 associated with the     cuts on the physical sheet
   \refs{ \EynardKG,\KazakovYH}.  
As a real manifold the  complex curve  represents a two-dimensional surface of genus 
 $g\le (p-1)(q-1)-1 $, with two punctures at $x=\infty$ and $y=\infty$.

  In the following we will consider the simplest  situation 
  when the  algebraic  curve has the topology of a sphere. 
  We will also  assume that 
\eqn\ppqq{ q=p+1,}
 which is sufficient  for describing  the  unitary 
 non-critical string theories.
Then  the function \CURVE\ has $(p-1)(q-1) -1$ double zeros, 
 and   the functions $X(y)$ and $Y(x)$ have  a single cut on their physical sheets.
  
  The projection of the complex curve to the $x$ complex plane, or the  
   Riemann surface of $Y(x)$,  represents  a  
  branched $p$-covering of the Riemann sphere and thus splits the complex curve into $p$ sheets.  
   The physical sheet  is the one that contains the puncture at  $x=\infty$.
      In addition, there are $p-2$ cuts  on the lower  sheets  
      that extends to infinity,
      where    the Riemann surface has  a branch-point of order $p-1$.   
    This branch point is the image of the puncture at $y=\infty$, which 
    we will denote by $x= \tilde \infty$. 
   We will denote  by $Y^{(k)}$, $k=1,..., p$,  the different branches of the function $Y(x)$  keeping the label  $k=1$ for the physical sheet (Fig.1, right).
       

 \bigskip
 
 \epsfxsize=300pt
\vskip 20pt
\centerline{
\epsfbox{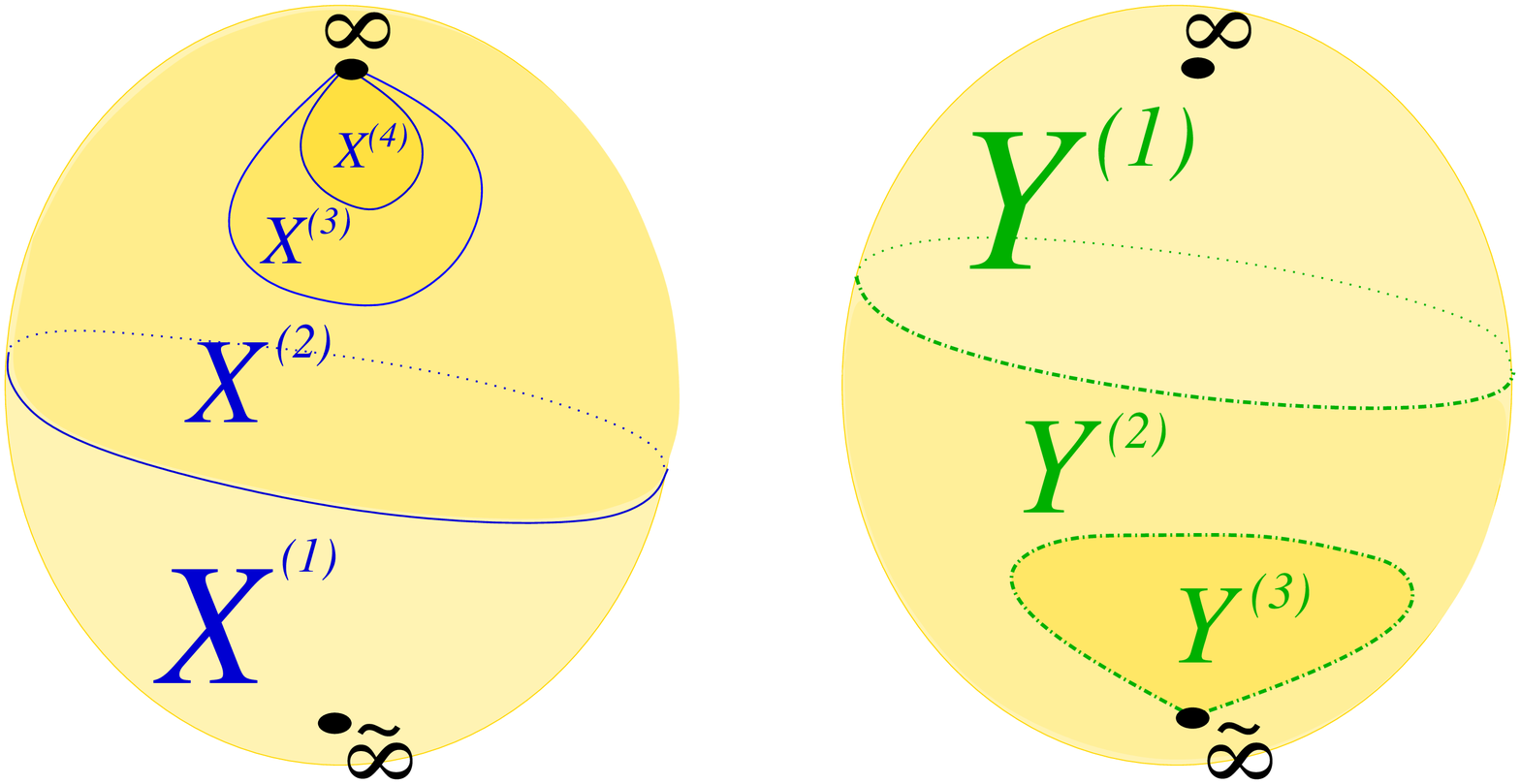 }
}
\vskip 5pt

 \bigskip

\vbox {\ninepoint  
\centerline{Fig.1 :  Sheets of the Riemann surfaces of $X(y)$
and $Y(x)$ 
 for  the one-cut solution}
 \centerline{ with  $p=3, q=4$.  
}
  }

 \vskip  15pt

\noindent
 In a similar way   the Riemann surface of $X(y)$ 
 splits the punctured sphere into  $q$  sheets $X^{(1)}, ...,X^{(q)}$
   (Fig.1, left).  Then the spectral curve can be rewritten as \EynardKG
  \eqn\spctrc{
  F(x,y)\sim  \prod_{k=1}^{p} \( y- Y^{(k)}(x)\)\sim
   \prod_{j=1}^{q} \( x- X^{(j)}(y)\).
   }

     We will refer to the points $x=\infty$ and $x=\tilde\infty$ 
as the north and south poles of the sphere.
The punctured sphere is characterized by the two cycles $A$ and $B$
dual to each other. The cycle $A$  goes along the equator   
 and the cycle
$B$ connects the  north and the south poles.
We assume that the sheets  can be  labeled  so that there is exactly one cut connecting  the $k$-th and the $k+1$-th sheet.
 The first and the last sheet  of the Riemann surfaces of $Y(x)$ and $X(y)$ have one cut and the other sheets  have two cuts.

  There is always possible to find an uniformization parameter $\o$
  that globally parametrizes the complex curve \DouglasPT,\DaulBG,
  \eqn\paramet{ X(\o) = \sum_{k=-1}^{q-1} X_k \,\o^k,
  \ \ Y(\o) = \sum_{j=-1}^{p-1} Y_k \, \o ^{-j}.
  }
One of the coefficients can  be  chosen arbitrarily because of the   symmetry 
with respect to rescalings of   $\w$.   The rest    $p+q+1$ coefficients  are determined
  through \ASYM\ as functions of the $p+q$ couplings $T_1,...,T_p, \tilde T_1, ...,\tilde T_q$ in \POTAB\  and the number $N$ of eigenvalues.
  The  punctured sphere is parametrized by the  complex plane $\o$. The north (south) pole of the sphere  then corresponds to the point 
  $\o=\infty$ ($\o = 0$).

\subsec{Perturbative  free energy  in terms of the spectral curve}
                                                                                              
\noindent
The    saddle-point equations \Sdpt\  mean that 
for some $k$ and $l$
\eqn\XYkl{x =
X^{(k)}(y), \ \ y= Y^{(l)}(x).  } Inverting the second equation we
find that for some $j\ne k$: $X^{(k)}(y)=X^{(j)}(y)$ which is
satisfied when $y$ is at one of the endpoints of the physical cut of
$X(y)$. The same is true for the variable $x$.  The ``perturbative''
saddle points are at the endpoints of the physical cuts of the
functions $X(y)$ and $Y(x)$
\foot{ The same algebraic curve can
describe several matrix models; they correspond to different real
sections that define different sets of local minima of the effective
potential \refs{\KazakovYH,\AKKnmm}.  Note that in the Normal matrix
model, which has the same complex curve as the Hermitian two-matrix
model, the saddle point is described by a closed contour and not by
isolated points.  This is possible because the Normal matrix model is
described by different real section of the complex curve
\TeodorescuQM.}.



 By the asymptotics  \ASYM, the integral  
 along the $A$-cycle is equal to the number of the eigenvalues:
\eqn\NORMD{  { N } =\oint_A  ydx   . }
 As was shown in \DavidZA (and generalized to the filling not only of
 the maxima but of of all extrema of the  potential
 \DijkgraafFC) for the one-matrix model and then for the two-matrix
 model in \KazakovYH, the integral of the function $Y(x)$ along the
 $B$-cycle is equal to the derivative in $N$ of the planar
 contribution $\CF_0$ to the free energy
\eqn\defFREN{
   \CF =  \log \CZ_N .
}
Indeed, the leading contribution  to the integral \CHAN\ is given by the 
saddle-point value of the effective potential \seff. Let   $x',y'$  be related by
$y'= Y^{(1)}(x') $ 
and $x'= X^{(1)}(y')$. There is always such a point along the cycle $B$.
 Then we write
\eqn\FPERT{ \eqalign{
\p_N \CF_0&=-\Sef(x', y')\cr &
= x'y' - \int _\infty^{y'} X^{(1)}(y)dy -
\int_\infty^{x'} Y^{(1)}(x) dx 
 \cr 
 &=  \int _\infty^{x'} Y^{(2)}(x)dx -\int _\infty^{x'} Y^{(1)}(x)dx \cr 
 &=\oint_B ydx .
}}

The formulas \NORMD\ and \FPERT\ determine the free energy in the
planar limit in terms of the two main cycles of the punctured sphere.
These formulas are geometrical in the sense that they do not depend on
the choice of the coordinate patches on the complex curve.  In the
next subsection we will obtain similar formulas for the
non-perturbative instanton contributions.

\vskip1cm

\subsec{ Leading order non-perturbative corrections}
  
\noindent 
Besides the  perturbative saddle point   
there are also other  saddle point solutions, which
describe the non-perturbative corrections to the free energy
\refs{\DavidSK,\DavidZA,\EynardSG}.  
 The meaning of the non-perturbative corrections depends on the
 physical context.  Typically they describe the decay of a metastable
 ground state caused by tunnelings of eigenvalues under a maximum of
 the effective potential (eigenvalue instantons).
    
    A comprehensive description of the instanton effects in the
    one-matrix model has been done in \DavidZA.  Once the effective
    action is known, one can generalize the analysis of \DavidZA\ to
    the case of the two-matrix model.  Here we will restrict ourselves
    to the leading non-perturbative corrections, which allow
    geometrical description in terms of the spectral curve.
    
      The  ``non-perturbative" saddle points  
  are double points of the complex curve represented by the pairs $x=x_{kl}, y=y_{kl}$ such that
 \eqn\SPEKJ{ y=Y^{(k)}(x) ,\ \ \ \ x=X^{(l)}(y) .  }
 At these points     $dF(x,y)=0$  and the effective action \seff\ has vanishing derivatives in $x$ and $y$.
 The  pairs $(1, l)$ and $(k,1)$ should be excluded because they 
 describe the same point of the curve (the second function is the inverse of the first).
 The pair $(2,2)$  should be excluded as well
  because it  determines  perturbative saddle point.
 The  number of the remaining   pairs  $(k,l)$ is 
 equal to the 
 maximal genus
 of the complex curve, $g_{\rm max}= 
 (p-1)(q-1)-1$.
   Among these  saddle points there are $(p-1)(q-1)/2  $   maxima
 and   $(p-1)(q-1)/2-1 $   minima of the effective action.
  In the last case the integration contour should be distorted in the
complex plane as explained, say, in Section 2 of \CallanPT.

  Each pair $(x_{kl}, y_{kl})$ 
correspond to two  {\it different} 
points of the complex curve  which  can also  be 
considered as a remnant of a collapsed handle, or vanishing cycle $A_{kl}$
of the complex curve in generic position \SeibergShih.
  Such a vanishing cycle is
associated with a pair of coinciding branch points of the Riemann
surface of $Y(x)$ or $X(y)$.
Let us denote by
  $\bar B_{kl}$   the corresponding $B$-cycle, which
  connects the 
   two poles of the sphere through   the pinched cycle $A_{kl}$.
Proceeding as in \FPERT\  we can write the effective
action at the $(k,l)$-th saddle point
\eqn\Smneff{ \eqalign{ S_{\rm eff} 
&= 
\int_{\bar B_{ kl}}  y \,dx 
=-\p_N\CF_0 + S_{kl}, }}
where
\eqn\Smn{S_{kl}=
\int_{B_{ kl}}  y \,dx .
} 
is the effective action associated with the compact cycle $B_{kl}$
 passing through the pinched cycle and the cycle $A$ (Fig.2).
  
   \bigskip
 
 \epsfxsize=135pt
\vskip 20pt
\centerline{ \epsfbox{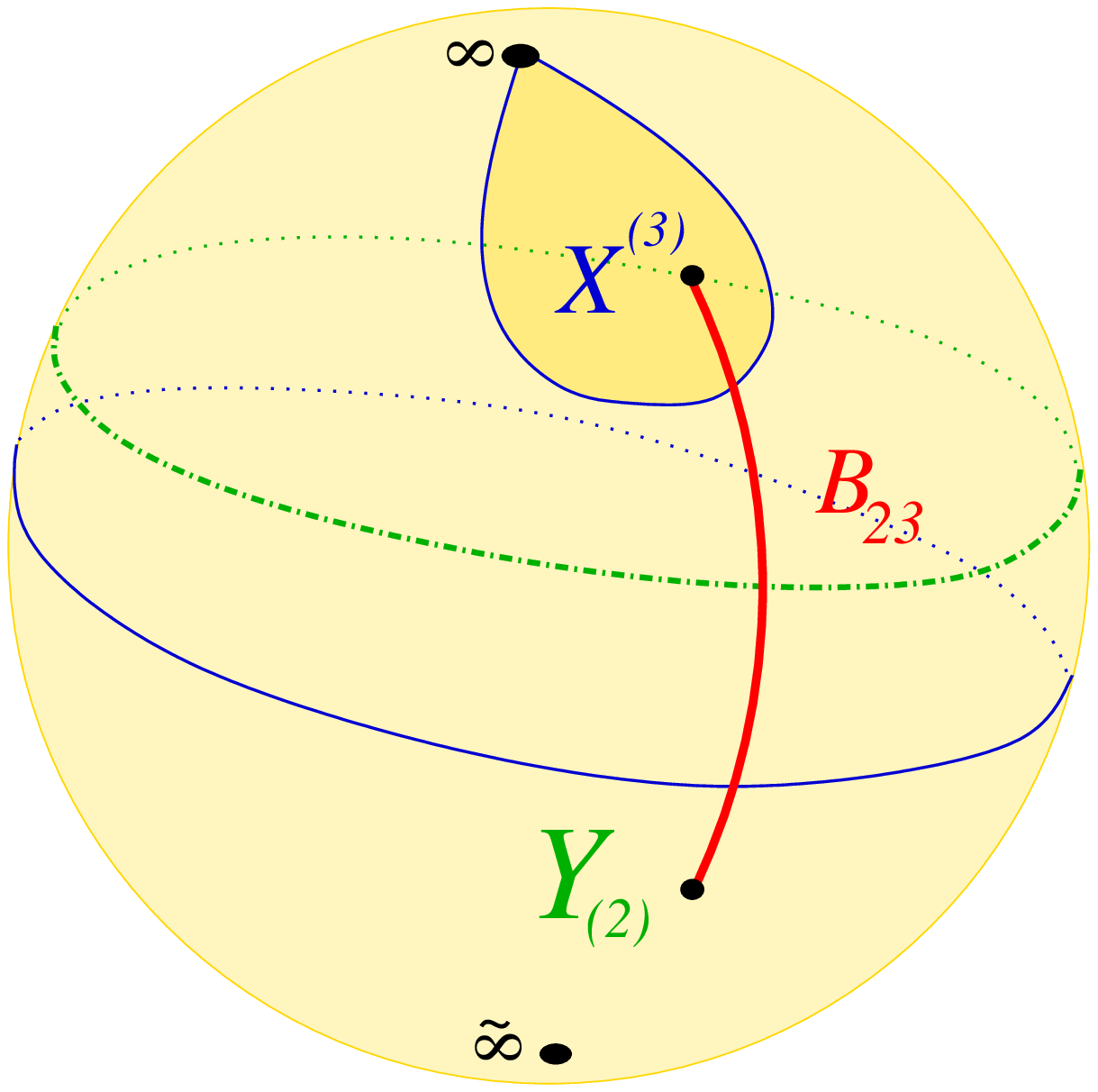 }
\hskip 30pt \epsfxsize=133pt \epsfbox{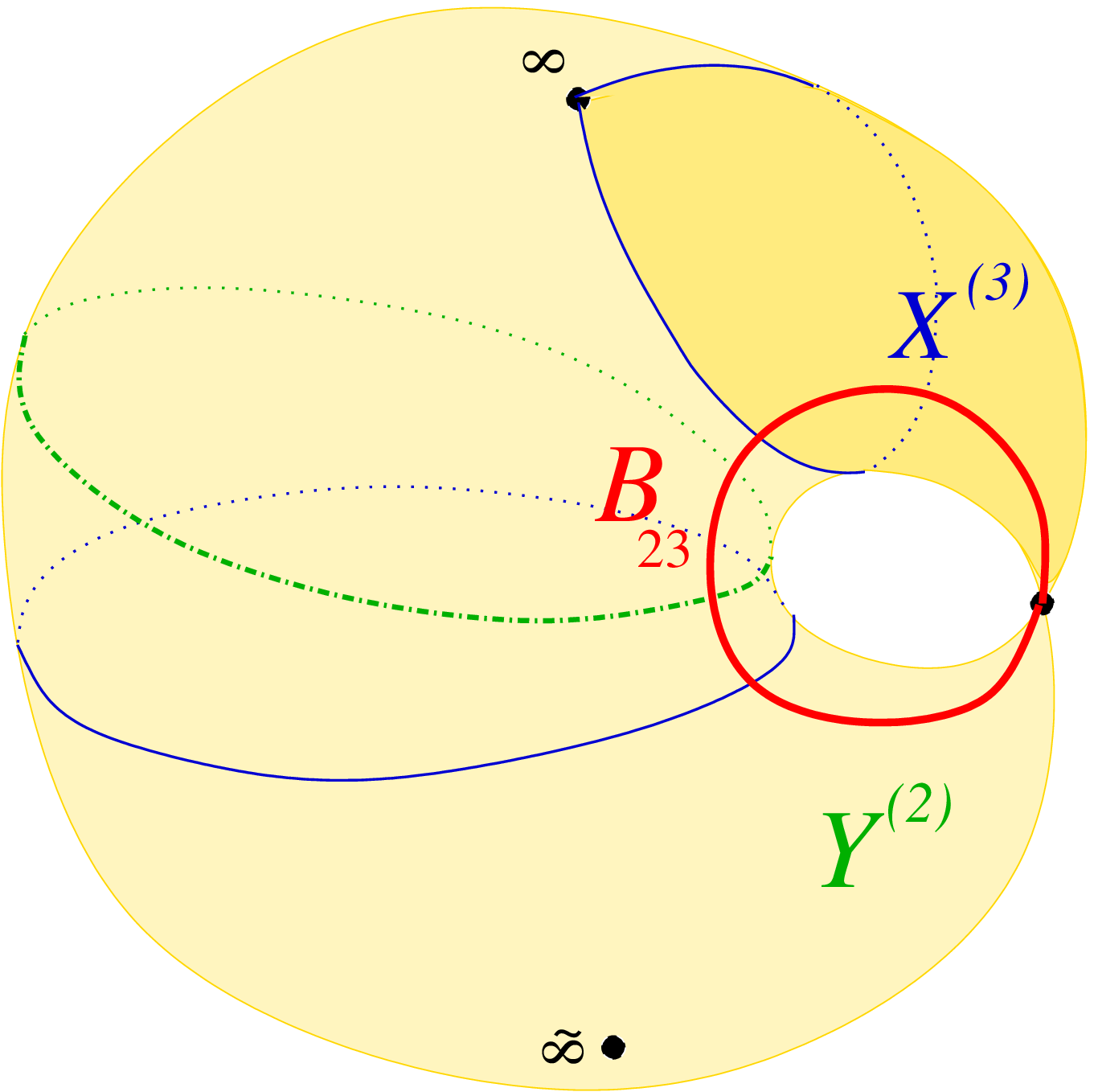 }}
\vskip 5pt

 \bigskip
{ 
 {\ninepoint 
\centerline{ Fig.2 :  The  only   cycle   $B_{23}$   for pure gravity ($p=2, q=3$).
 It   can be viewed either }
\centerline{  as an open contour  connecting 
the two double points on the sphere (left) or }
\centerline{ as a closed contour 
going through a pinched cycle of a torus (right).
 }
 }
 \vskip  25pt

  \noindent
If we take into account both minimal and non-minimal saddle points,
eqs. \CHAN\ and \Smneff\  give
\eqn\chann{  \eqalign{ 
{\CZ_{N+1}\over \CZ_N}&\simeq
 e^{ \p_N\CF_0}
\(1+ \sum_{k,l}
  c_{kl} \, e^{- S_{kl}}\) }} 
where the sum is taken over all pairs $(k,l)$ discussed above.
  Written
for the free energy $\CF=\log \CZ_N$, this formula reads
 \eqn\chanF{
 \p_N \CF= \p_N\CF_0+  \sum _{(k,l)\ne (2,2) }  c_{kl}\,e^{-  S_{kl}}
 +...}
As was explained in \refs{\DavidSK,\DavidZA}, the non-perturbative
corrections should be understood not as an improvement of the $1/N$
expansion but rather as a difference between the two free energies
corresponding to the two ways of distortion of the integration
contours in the matrix integral into the complex plane.

We believe that in any matrix model described by algebraic curve the
non-perturbative corrections are given in the planar approximation by
the formula \chanF\foot{We have checked that \chanF\ holds also for
the $ADE$ matrix chains defined in \adem.}. 
 In the next section we will apply \chanF\ to the critical regimes of
the 2MM, in which case it reduces to the similar formula obtained in
\SeibergShih.

\newsec{Explicit results for the $(p, p+1)$ critical points}


\subsec{The scaling limit near the  $(p,q=p+1)$ critical point}

 \noindent 
 The  noncritical string theories are described by the critical points  of the 
 complex curve. A critical point can occur when a branch point comes close to a double point or to another branch point. 
 Here we will examine the ``maximal'' critical point that arises when  the 
 right   branch point  point of the physical cut  of $Y(x)$  coalesce with the $p-2$   branch 
 points on the lower sheets.  
    At this   point  the   equation of the  curve \paramet \  
  takes the form    \DaulBG
 \eqn\critpar{
 X(\o)=N_c\  {(1-\o)^p\over \o}, \quad Y(\o) = N _c\ \o {\(1-{1\over \o}\)^q}.}
 In the vicinity of   the origin
 the complex curve degenerates to  
  \eqn\scalingpq{ y^p =  x^{q}.}
The solution \critpar\ corresponds to certain choice of the coupling constants
 in \POTAB, all of order of $N_c$.    One can  introduce a one-parameter 
 deformation  of this singularity by
  changing the number of eigenvalues to
 $N<N_c$.    The difference $N_c-N$ is proportional to the cosmological constant 
 $\mu$ of the corresponding $(p,q)$ string theory.
  By introducing a small  cutoff parameter $a$
   and  rescale the  variables as
  \eqn\rescal{ {x\over a^{q} }\to x, \ {y\over a^p}\to
 y,\ \ {N -N_c\over  a^{p+q} N_c}\to
     - {\mu} , \ \ a^{p+q} N_c \to {1\over g_s}} 
   we blow up the
the vicinity of  the (deformed) critical point so that 
all the cuts  become semi-infinite  (Fig.3).

 \bigskip
 
 \epsfxsize=350pt
\vskip 20pt
\epsfbox{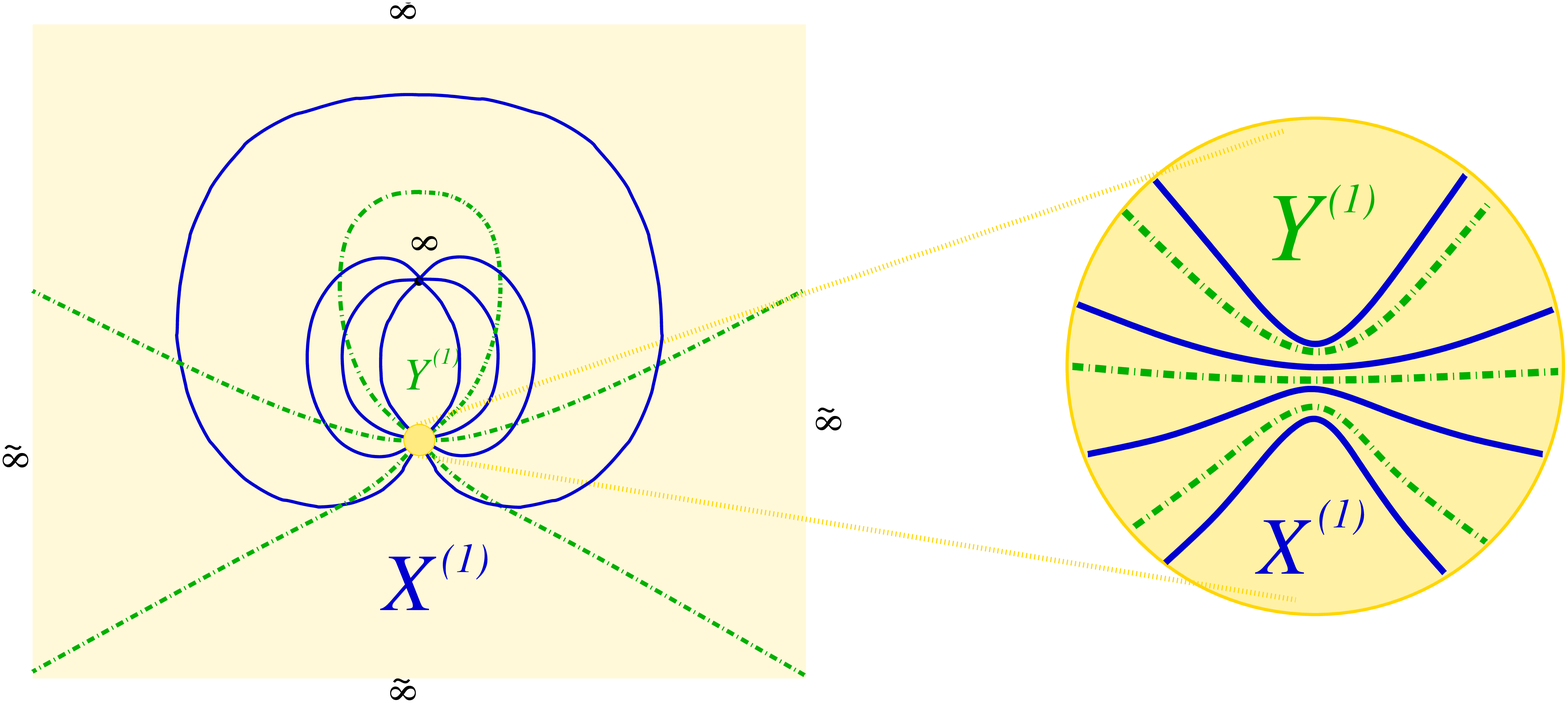 }
\vskip 5pt

 \bigskip

 \bigskip

{\ninepoint \noindent
\centerline{Fig.3 :   Sheets   near  the $(p,q)$ critical point 
with $p=3$, $q=4$,  in the }
   \centerline{parametrization \critpar.  On the right, the blown up scaling domain.   }
 }
 \vskip  15pt

\noindent 
The physical cuts of the functions $y=Y(x)$ and $x=X(y)$
  extend  to infinity  along the intervals $-\infty<x<-2M$ and
     $-\infty<y<-2\tilde M$ respectively, 
  where $M\sim \mu^{1/2}$ and $\tilde M\sim  M^{q/2p}$.  
  We will normalize $X$ and $Y$ so that  $M$, $\tilde M$ and $\mu$ are related by
 \eqn\Mxi{
 M= \xi^p, \quad \tilde M = \xi^{q}, \quad \mu = \xi^{2p}.
 } 
A very useful parametrization in the scaling limit is given by
expanding $X$ and $Y$ in the Chebyshev polynomials of a third variable
$$z= 2\xi \cosh\th,$$
 which appears naturally in the formalism using the dispersionless KP hierarchy \DaulBG :
\eqn\CHEB{\eqalign{ x &=  2T_p(z/2)=2\xi^p \cosh p\th, \cr 
y &=2 T_q(z/2) =2\xi^{q} \cosh q \th.}}
To avoid the subtleties that arise in non-unitary theories, we will assume
below  that $p=q+1$.
The parametrization \CHEB\ unfolds the branch points of the functions
$Y(x)$ and $X(y)$ and allows us to work with entire functions of
$\th$.  The sheet structure is shown in Fig.4.
The $k$-th sheet of the function $Y(x)$ is parametrized by the semi-infinite strip
$$\Re\th> 0,  \qquad   {k\over p}\pi  <|\Im\th | <  {k-1\over p}\pi $$ 
and the $l$-th sheet of  the function $X(y)$ is parametrized by the semi-infinite strip
$$\Re\th> 0, \qquad     {l\over p+1}\pi  <|\Im(\pi - \th )| <  {l-1\over p+1}\pi .$$
 At the critical point the parameter  $z$  is related to the 
parameter $\o$ in \critpar\ by 
$$ \w = 1 + a z.$$

%
%
 
 \epsfxsize=200pt
\vskip 20pt
\hskip 80pt
\epsfbox{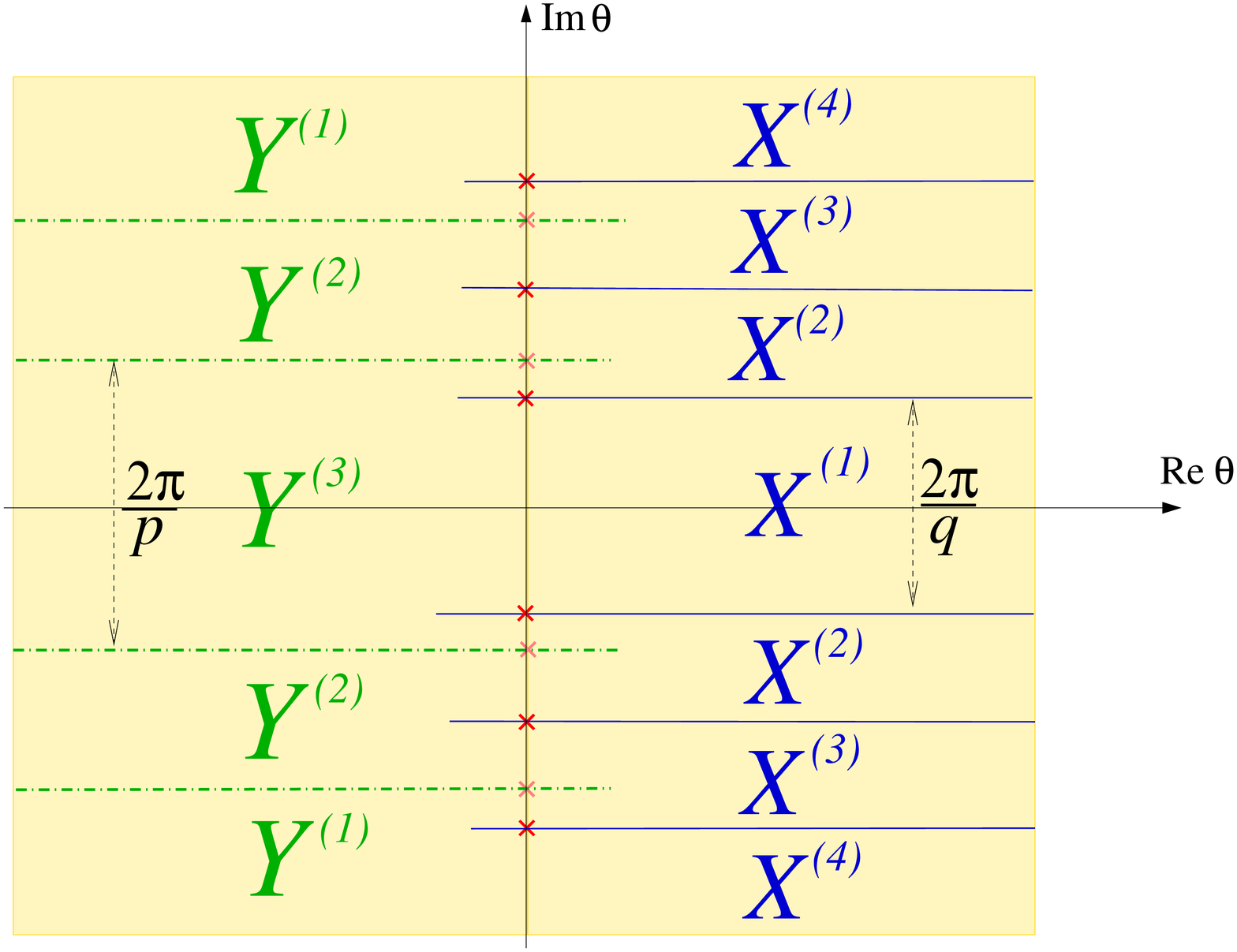 }
\vskip 5pt

 \bigskip

 \bigskip

{\ninepoint \noindent
\centerline{Fig.4 :   Sheets   in the $\th$ parametrization 
for  $p=3$, $q=4$. }
\centerline{All functions are symmetric under reflection $\th\to -\th$.}
   \vskip  10pt

 In any $(p,p+1)$ theory of 2D gravity with matter central
charge $c=1-{6\over p(p+1)}$, the function $Y(x)$ defined by \CHEB\
gives the loop amplitude, that is the disc amplitude with a marked
point on the boundary, with the boundary cosmological constant $x$ and
bulk cosmological constant $\mu$. The function $X(y)$ does the same
for the dual $(p+1,p)$ theory\foot{The duality among $(p,q)$ and
$(q,p)$ theories was first observed in the matrix models in 
\refs{\KharchevKD,\KharchevAS}}.  The expression \CHEB\ has been
first derived in \KostovHN\ in the context of the $ADE$ loop gas
models.  Its interpretation from the point of view of Liouville theory
is given in \MooreIR.

 The two potentials \AVLOG\ can be calculated in the scaling limit by
 integrating \CHEB. This gives $(q=p+1)$
  \eqn\Phivsc{\eqalign{ \Phi&= {2p \ \xi^{p+q } } \left( {
  \cosh(p+q)\th \over p+q} -{\cosh (q-p)\th\over q-p} \right)\cr
  \tilde \Phi& = {2q \ \xi^{p+q }  } \left( { \cosh(p+q)\th \over
  p+q} +{\cosh (q-p)\th\over q-p} \right)} }
  and one can check that  the effective action is constant 
  on the complex curve (equal to zero 
  in our  conventions):
 \eqn\Legndre{ \Phi(x)+\tilde\Phi(y) =xy .}  

 To make connection with
 the boundary Liouville theory we introduce the parameters
\eqn\defbs{ b = \sqrt{p/q}, \qquad \pi b s = \sqrt{pq}\ \th.
}
Then the equation of the complex curve  becomes
\eqn\CHEBL{ x = 2\mu^{1/2}\cosh \pi b s, \qquad y = 2\mu^{1/2b^2} \cosh
\pi s/b.  }


\subsec{Perturbations   by relevant operators  around the $(p, q=p+1)$ critical point}

\noindent
A  generic  perturbation of the  $(p,p+1)$ critical point  
 is described by a curve
that  behaves as   \scalingpq \  at infinity. 
 Such a curve  has the parametric  form 
%
\eqn\POLC{\eqalign{
x&= x(z)= z^{p}+ x_{p-1} z^{p-1}+ x_{p-2}
z^{p-2}+\ldots+y_0
,\cr
 y&=  y(z)=  z^{p+1}+ y_{p-1} z^{p-1}+ y_{p-2}
z^{p-2}+\ldots+y_0
.}
} 
The      curve     \POLC\       has  only one singular point at
infinity where the function $y(x)$ has a branch point of order $p$ and
the function $x(y)$ has a branch point of order $q=p+1$.

The original Toda integrable structure  of the two-matrix model is
characterized by two singular points,
$\infty$ and $\tilde \infty$, and the  spectral curve is determined by 
   two asymptotic conditions \ASYM\  associated 
    with the two punctures. 
For  degenerate curves of the 
form     \POLC, the  relevant integrable structure is that of 
      the $p$-reduced dispersionless 
      KP hierarchy \KricheverQE  , which has 
 only one singular point $y=x=\infty$. 
       Therefore
in the scaling limit it is sufficient to introduce a single local
coordinate in the neighborhood of the puncture, say $x$,
 and the
corresponding set of couplings $t=\{t_n\}_{n=1}^{\infty}$ 
 defined by the coefficients of the powers $x^{n/p-1}$, $n=1,2,...$,
 in the Laurent series of $y(x)$ at infinity:
\eqn\timestn{ y(x) \equiv \p_x\Phi 
=   {1\over p}  \sum_{n\ge 1 }  \(
n \  { t_n } \ x^{n/p-1}
 \ + \  v_n  \  x^{ -n/p - 1}\).  } 
The  curve \POLC\ is described in terms of the  non-zero
couplings $t_1,...,t_{p-1}, t_{p+1}, ..., t_{2p+1}$.   
 The coefficients   $v_n$  in   \timestn\  are functions  of these  couplings.
    The     integer powers in $x$ do not have  a branch point at infinity and
  therefore the sum is restricted to $n\ne 0 \ ({\rm mod } \ p)$.
 
 The integrable perturbations   associated with the  couplings 
$ t_n$ 
are generated by the Hamiltonian flows of the $p$-reduced KP hierarchy.
Following Krichever     \KricheverQE,   one can associate with  the
 Hamiltonian   coupled to $t_n$
  a meromorphic differential $dH_n (x, t)$.
   The form of the classical Hamiltonians are determined by  
  the complex curve   \POLC. 
  The  differential of the effective potential  $\Phi(x, t)$  is  given by
 %
 
 \eqn\dPhi{\eqalign{
 d\Phi &= ydx + 
  \sum_{m\ge 1 }  H_{m}  (x) d  t_{m } 
  .}  } 
The   dispersionless KP hierarchy  can be interpreted as a 
 classical   Hamiltonian system   with  Poisson bracket 
\eqn\poissonb{
\{ f,g\} =  {\p f\over \p z} {\p g \over \p  t_1} - {\p f\over \p t_1} {\p g \over \p  z}.
}
The differential \dPhi\ then can be considered as a generating function for the 
canonical transformation between the phase-space coordinates $(z,  t_1)$ and $(x,y)$.  
Given the curve \POLC\ and the expansion \timestn\ at infinity, the
   expressions for the classical Hamiltonians are
  \eqn\Hnofom{ H_m =  [ X^{m/p}]_+}
 where $[\ \ ]_+$ denotes the non-negative part of the Laurent series
of the function $X(z )$ at infinity. 
  It follows from   \dPhi\ that the  parameter  
    $z $    in the definition of the curve  \POLC\ 
    can be identified 
     with the first Hamiltonian $H_1$:  
    \eqn\Honwo{
    H_1 =   \ [ X^{1/p}]_+=   z .}
   It is technically convenient to  use the  parametrization \POLC\ 
 and
  consider  
 $z  = H_1$  as a  global coordinate on the  complex curve.
 From   \dPhi\ and \Hnofom\  one finds the expression of
 the effective potential 
 \eqn\Phigen{\eqalign{
 \Phi  =
  \sum_{m\ge 1}  H_{m} (z , t)  \  t_{m } 
  .}  } 
Given the relevant couplings $t_1,...,t_{p-1}, t_{p+1}, ...,
  t_{2p-1}$, from \Phigen\ and \Hnofom\ one can then  evaluate the $2p-2$
  coefficients defining the curve \POLC.

Now let us   return to the special case of 
 the curve \CHEB. 
 The only deformation parameter here, 
 the cosmological constant $\mu$, 
  can be identified, up to a normalization, with the 
 coupling $t_1$. 
  Plugging \CHEB\ into   \Hnofom\ one finds for the classical 
 Hamiltonians  \DaulBG
\smallskip
 \eqn\paraH{
 \eqalign{
H_m&=   2  \xi^m\, \cosh m\th,    \qquad m=1,2,...., 2p-1;\cr
& \cr
H_{2p+1} &=  2 \xi^{2p+1} \(  \cosh (2p+1)\th +  {2p+1\over p} \, \cosh\th \) .
 }
}
Comparing with the expression 
\Phigen, we conclude that   the curve \CHEB\ 
 describes  the point 
 \eqn\bcgrnd{
 t_{2p+1}  = {p\over 2p+1},\ \ \ \ \ 
t_1 = -(p+1) \,\mu, \ \ \ \ \ t_{\rm others} =0}
  in the space of couplings. 
 The    relation   between $\mu=\xi^{2p} $ and $t_1$
  follows also directly from the  
  classical string equation  $\{x,y\}=1$   applied to  the solution  \CHEB.
  

%
%

\vskip1cm
  
 \subsec{  Free energy and two-point functions}

 \noindent    
Here we will calculate   the 
two-point functions of relevant operators on the sphere for the
background \CHEB.  
  The pair of dual variables $\mu$ and $\p_\mu \CF$ are  expressed in
terms of the complex curve through the formulas \FPERT\ and \NORMD. In
terms of the rescaled variables defined by \rescal\ these formulas
read
\eqn\FPERTs{ - \p_\mu\CF= \oint_{B} ydx = \quad \oint_B d\Phi } 
and 
\eqn\NORMAs{ -\mu= \oint_{A} ydx= \quad \oint_A d\Phi  . }

  The general
formulas for the 2MM with polynomial potentials, of the type obtained
in \BertolaKZ\ cannot be immediately applied for this task in the
critical limit.    
 In the scaling limit the integrals in 
 \FPERTs\ and \NORMAs\ diverge and need regularization.  One could do it
 explicitly by making the cut finite.
 It is however possible to extract the necessary information from 
  \FPERTs\ and \NORMAs\   without any regularization by using the
   {Riemann bilinear identities} (RBI).
RBI follow from the fact that there are no meromorphic $(2,0)$-forms
on one-dimensional complex curves.  For each pair of holomorphic
differentials $d\Omega_1$ and $d\Omega_2$,
\eqn\RBI{ \eqalign{ 
0&=\int_{\Sigma} d\Omega_1\wedge d\Omega_2 \cr 
&= 
 \oint_{A}d\Omega_1\oint_{B} 
d\Omega_2-\oint_{A} 
d\Omega_2\oint_{B}d\Omega_1 \  - 
\sum {\rm res} \left(d\Omega_1 \Omega_2\right) .
}} 
  In particular, when
\eqn\doma{
d\Omega_1=\p_\mu Y(x)\,dx= d\p_\mu\Phi,
 \quad d\Omega_2=\p_m\p_n
Y(x) \, dx=d\p_m\p_n \Phi   ,}
  equations  \FPERTs\ and \NORMAs \ yield
\eqn\RICR{  \eqalign{ 
{1\over 2\pi i}\oint 
d\p_\mu\Phi (x) \cdot\p_m\p_n \Phi(x) 
=\p_\mu\p_m\p_n\CF .}} 

A rigorous derivation of  this identity can be  done using the
 the fact that the partition function of the two-matrix model is a  
 $\tau$-functions for the KP integrable hierarchy, which means that
 the  coefficients  $v_n$ in \timestn\ are related to the free energy by 
$ v_n = \p_n\CF.$
This implies   the following identities, which have been derived in  
\refs{\KricheverQE} (for more explicit derivation see
    \KharchevCP,\ChekhovGZ)
\eqn\ONEPF{ \eqalign{    \p_n\CF&=- 
{1\over 2\pi i}\oint _\infty x^{n/p}   d\Phi
,\cr
\p_m\p_n\CF& = -
{1\over 2\pi i }\oint _\infty x^{n/p}    dH_m
\cr
\p_k\p_m\p_n\CF&
=- {1\over 2\pi i}\oint _\infty x^{n/p} 
d \p_k H_m
={1\over 2\pi i}\oint _\infty 
\(\p_kH_m\)_x   dH_n.
}}
 Since $ \mu\sim t_1$, 
 equation  \RICR\ is a particular case of the last of these relations.
 
 Our aim is to calculate the two-point   function
 $\p_m\p_n\CF$ in    the classical
   background  \CHEB, where the classical  Hamiltonians are 
   given by \paraH.   Inserting 
%
\eqn\KYXH{ \(\p_\mu H_m\)_x=  
  m \xi^{-2p+m} {\sinh (p-m)\th \over p \sinh p\th} 
, \ \ 
\ \ dH_n=  2n \xi^{n}   \,
{\sinh n\th } \, d \th
}
%
%
  into \ONEPF\  we get 
\eqn\DMDKY{  \p_\mu \p_m\p_n\CF = {2 mn }\
\xi^{m+n-2 p}\  
 \oint {d\th\over 2\pi i}\  {\sinh n\th\ \sinh(p-m)\th \over p\sinh 
 p\th} .} 
We evaluate the integral  along the  contour $\Im \th \in[0,2\pi)$, $\Re\th\to\infty$,
which gives
\eqn\FINF{ \p_\mu\p_m\p_n\CF
= 
{mn\over p }\ 
\xi^{2(n-p)}\ \delta_{m,n}      .
} 
Integrating with respect to  $ \mu = \xi^{2p}$
 we find for the two-point function  
\eqn\FINKF{    \p_n^2\CF=  
{n}\xi^{{2n}}  .} 
  For $k=1$ we obtain from here for the string susceptibility 
\eqn\susc{ \p_1^2\CF 
=
\xi^{2}. }

\vskip1cm

\subsec{ Eigenvalue  instantons  at the $(p,p+1)$ critical point   }

%
%
\noindent 
As was  discussed  in section 2.3,  the eigenvalue instantons 
are associated with the  the  points 
  \SPEKJ\ of the curve where conical singularities occur. 
  These points are given by the non-trivial solutions
$(\th \ne \th')$ of the equations
\eqn\CLOSE{  x(\th )=x(\th '),\ \ \ \ \   y(\th )=y(\th '),}
where the functions $x(\theta)$ and $y(\theta)$ are defined by
\CHEB. The solutions of \CLOSE\ are $ \th = \th _{mn} ,
\
\th ' = \th _{m,-n}$ $(m=1,..., p-1; n=1,..., p)$ with
\eqn\defsmn{
\th _{mn}= i\pi \({m\over p} +{n\over p+1}\).}
The equations \CLOSE\  can be written also in the form \SPEKJ,
with $k= [|m-n {p\over p+1} | ]$ and $l = [|m{p+1\over p} +n| ]$.
Due to the symmetries   $\th\to -  \th$ and
$\th\to \th + 2\pi  $,  
  the points with parameters $\th_{mn}$ and $\th_{p-m, p+1-n}$
describe the same double point of the complex curve.
Therefore  in the scaling limit there are  only
${(p-1)(q-1)\over 2}$ double points 
 which correspond to 
local  maxima of the effective action. The ${(p-1)(q-1)\over 2}-1$ 
double points that correspond to 
local minima  are sent to infinity after the blow-up.
In Fig.5 we give the  plots of the  closed contours 
in the $(x,y)$-plane 
for   $p=2,3,4$. Some of the cycles have backtracking parts, which 
can be deleted.  The effective action associated with each cycle is 
 proportional  to its (algebraic) area.

  
  \vskip 28pt

  {\ninepoint 
  \hskip 2pt
  \centerline{  
 \vbox{
      \hbox{
            \vbox{ \hbox{  \epsfxsize=50pt\epsfbox{ 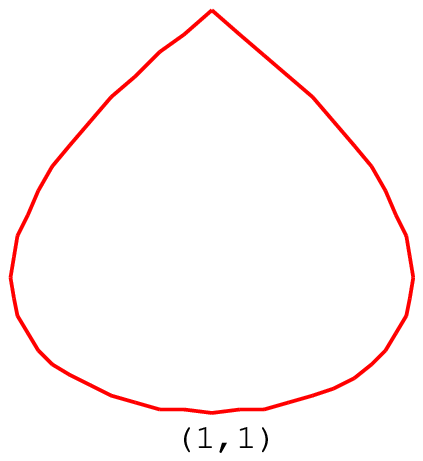 }}
                        \hbox{ Pure gravity $(p=2)$}}
   \hskip 1cm\
             \vbox{
          \hbox{
                \epsfxsize=50pt\epsfbox{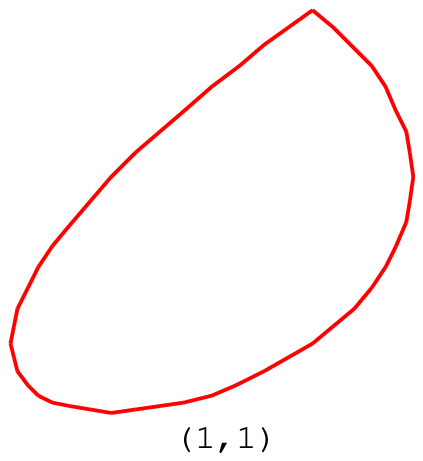}
                 \epsfxsize=50pt\epsfbox{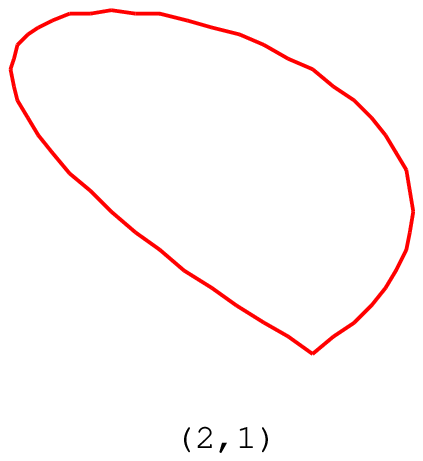}
                   \epsfxsize=50pt\epsfbox{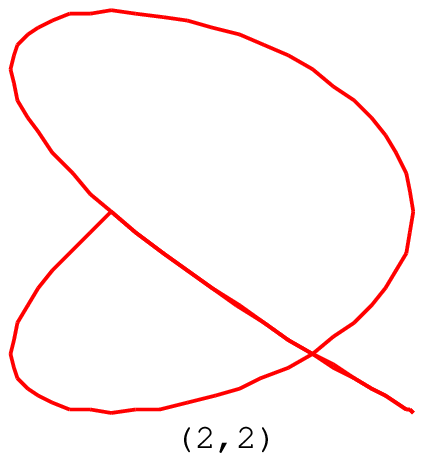}}
      \hbox{\qquad Ising model coupled to gravity  $(p=3)$} 
      }}
      \vskip 1cm
   \hbox{  \vbox{
     \hbox{ 
      \epsfxsize=50pt\epsfbox{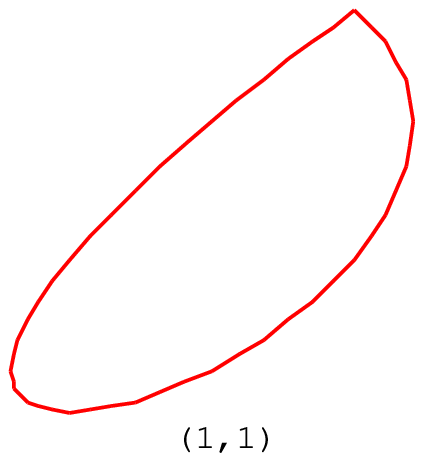}
    \epsfxsize=50pt\epsfbox{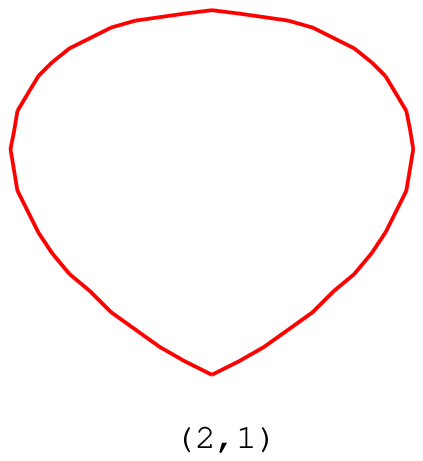}
      \epsfxsize=50pt\epsfbox{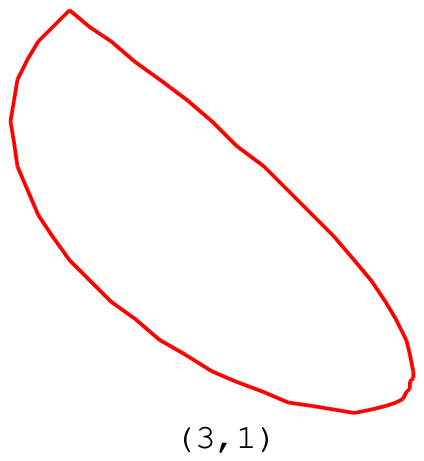}
      \epsfxsize=50pt\epsfbox{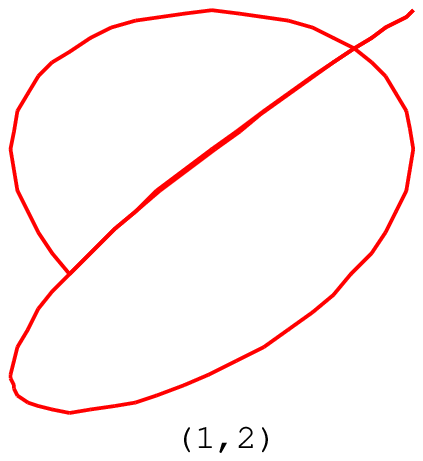}
    \epsfxsize=50pt\epsfbox{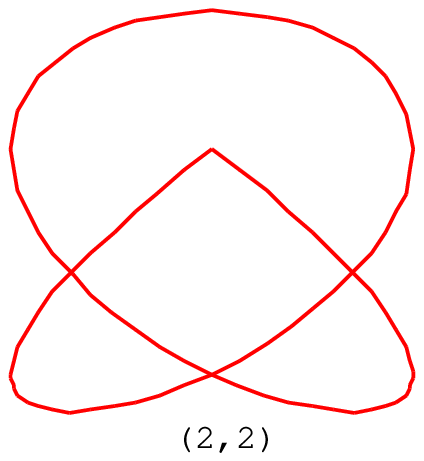}
          \epsfxsize=50pt\epsfbox{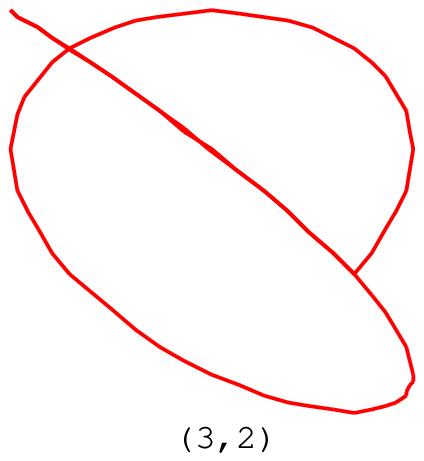}
      }
           \hbox{ \hskip 2cm Tricritical Ising  model  coupled to gravity  ($p=4$)}
       }}
       }}

         \vskip  10pt

          \centerline{Fig.5:    
          The  instanton  cycles for the  $(p,p+1)$ non-critical strings 
           with p=2,3,4. }
}
      

      \vskip  26pt

\noindent
 The instanton contribution to the effective action corresponds to
  the non-trivial cycle $B_{mn}$ and is given by \Smn
\eqn\Ssmn{
S_{mn}=  \int_{B_{mn}} Y(x) dx =
\Phi(\th_{mn})-\Phi(\th_{m, -n}).  } 
 Evaluating the difference with $\Phi$ given by \Phivsc, we get

\eqn\INSS{
S_{mn}=    {8 p (p+1)\over 2p+1} \xi^{2p+1} 
 \sin( \pi m /p) \ \sin ( \pi n/q).  }

We can also calculate the first order 
corrections\foot{The pre-factors $c_{kl}$ in \chann\ 
  were conjectured in 
\AlexandrovNN\   to  behave as   $c_{mn}\sim
  g_s^{1/2}$ for all $c<1$ theories, as is known from some explicit
  examples (see
\refs{\GinspargCY,\AlexandrovNN}). } 
 $\delta S_{mn} = \sum _k t_k \p_{k} S_{mn}$
to the instanton  effective action  in presence of perturbations
by the order operators $\CO_k$,  $ k=1,...,
p-1$:
\eqn\INSTK{
\p_{k} S_{mn}   =H_k(\th_{m, n})- H_k(\th_{m,-n})
=-{4} \xi^{k}\ 
\sin {\pi kn\over p}\sin {\pi km\over q}   .}

To compare with the world sheet CFT we need to prepare a dimensionless
ratio that is not sensible to the normalization of $x, y$ and $\mu$.
We introduce the dimensionless ratio of \INSTK\ and \FINKF
\eqn\RATIO{  r^{(k)}_{m,n}
= {  \p_k S_{mn}\over \sqrt{\p_k^2\CF_0 }}= -
{4 \over\sqrt{ k} }\sin {\pi kn\over p}\sin {\pi km\over q}   }
For $k=1$ the result coincides with those found in this particular
case in \EynardSG\ and reproduced within the Liouville CFT approach in
\AlexandrovNN.  We will show now that this result can be reproduced
also for any $k$ from the CFT.

\newsec{One-point functions on ZZ branes and instantons}

\noindent
Let us compare the result \RATIO\ to the normalized one-point
functions on ZZ branes in a Liouville theory with 
$Q=  \sqrt{p+1\over p}+\sqrt{p\over p+1}$, in the same spirit as it was done in
\refs{\MartinecKA,\AlexandrovNN}. The  Hamiltonians $H_k$ 
are described in the CFT approach by  the product of matter and Liouville operators
\eqn\mapMMCFT{
H_k \quad \to\quad \int \limits_{\rm Disc} \CO_k \cdot e^{2\a_k\phi}}
We assume the most general boundary conditions for the matter 
and Liouville fields. The matter boundary conditions are labeled by the 
entries of the Kac table, which we denote by $(m,n)\sim (p-m,p+1-n)$ with 
$1\le m\le p-1$ and $1\le m \le p$ \CardyIR. The ZZ boundary conditions for the Liouville field
  $(m',n')$ with $m',n'\ge 1$  \ZamolodchikovAH.
We will need the following formulas:

 \smallskip
    \smallskip
\noindent
--  The structure constant for the
disc one point function of
the  matter order operator $\CO_k$
in presence of the boundary condition
 $m,n$ 
\refs{\BehrendBN,   
  \RecknagelRI}:
\eqn\CARDY{  \< \CO_{k} \>_{m,n}^{^{\rm matter}}=\({8\over p(p+1)}\)^{1/4}
{\sin {\pi m k\over p}\sin{\pi n k\over p+1} \over 
\sqrt{\sin {\pi k\over p}\sin{\pi k\over p+1}}}  .}

 \smallskip
    \smallskip
\noindent
-- The disc partition function in presence of  $m',n'$  ZZ-boundary condition 
of the Liouville vertex operator
$e^{2\a_k\phi}$, $\a_k = \hf\(Q- {k\over \sqrt{p(p+1)}}\)$
  \ZamolodchikovAH:
\eqn\ZZB{  \<  e^{ \a_k \phi }\>_{m',n'} ^{^{\rm Liouv}}
=  \<  e^{  \a_k \phi } \>_{1,1}^{^{\rm Liouv}}\ \ 
{\sin { \pi k m' \over p}\sin {\pi k n'\over p+1} \over 
\sin {\pi  k\over p}\sin {\pi  k\over p+1} }
}
where\foot{See the discussion on the origin of the first factor 
   at the end of subsection 2.2 of the paper 
\AlexandrovNN.}
\eqn\WAVEF{  \eqalign{
  \<  e^{  \a_k \phi } \>_{1,1}^{^{\rm Liouv}}\
=-  \sqrt{2\over \pi}\   M^{k/p}   { 2^{3/4} \pi k    
\over\sqrt{p(p+1)} \Gamma\(1+{k\over p+1}\)\Gamma\(1+{k\over p}\)}
}} 
and the constant $M$  is  related to the 
  Liouville  bulk cosmological constant  $\mu_L$ by
  \eqn\Mmol{
 M~= ~\sqrt{ \pi \mu_L \g\({p\over p+1}\)},\qquad
 \gamma(x)\equiv{\Gamma(x)\over\Gamma(1-x)}~.}
  
  \smallskip
    \smallskip
  \noindent
  --The Liouville two-point function on the sphere  
\refs{\DornXN,\ZamolodchikovAA}
(see the Appendix and the eq. (3.45) of \AlexandrovNN\ for this
particular formula):
\eqn{\TWOPF}{\eqalign{
\< e^{ \a_k \phi } e^{ \a_k \phi }\>^{^{\rm Liouv}}_{\rm sphere}
=-{k\sqrt{p(p+1)}\over 2\pi p^2  }
M^{2 k/ p}\ 
& \gamma\( 1-{k\over p+1}\) \gamma\(- {k\over p}\).
}}

Now we have to see which  matter and Liouville boundary conditions 
can reproduce the  dimensionless  ratio \RATIO\ obtained from the two-matrix model. 
 The explicit formula for $r^{(k)}$ obtained by combining
 \CARDY, \ZZB \ and \WAVEF\ is

\eqn\RRSO{  r^{(k)}_{m,n;m',m'}={ \< \CO_{k} \>_{m,n}^{^{\rm matter}}
 \<  e^{  \a_k \phi } \>_{m',n'}^{^{\rm Liouv}}
\over \sqrt{ \< e^{ \a_k \phi } e^{ \a_k \phi }\>^{^{\rm Liouv}}_{\rm sphere}
}  }=\rho_k \sin {\pi m k\over p}\sin{\pi n k\over p+1} \sin {\pi m' k\over p} \sin {\pi n' k\over p+1}
}
with 
\eqn\RHOO{\eqalign{ \rho_k &={  \< \phi_{k,k} \>_{1,1}^{^{\rm Matter}} 
   \cdot 
 \<  e^{  \a_k \phi } \>_{1,1}^{^{\rm Liouv}}
 \over \sqrt{ \< e^{ \a_k \phi } e^{ \a_k \phi }\>^{^{\rm Liouv}}_{\rm sphere}
} }\cr
&= \sqrt{2/\pi}{\({8\over p(p+1)}\)^{1/4}
\(\sin {\pi k\over p}\sin{\pi k\over p+1}\)^{-1/2}
 2^{3/4} \pi {1\over \sqrt{p(p+1)}} k \over \Gamma\(1+{k\over
 p+1}\)\Gamma\(1+k/p\) \[-{k\sqrt{p(p+1)}\over \pi p^2 }\gamma\(
 1-{k\over p+1}\)\gamma\(-k/p\)\]^{1/2}} \cr 
&=-{4\over \sqrt{k}} 
.}}
We see that  the matrix model result  matches with a particular subset of  
boundary conditions.   The exponent of the   tunneling amplitude in the matrix model
associated with the cycle $m,n$   is reproduced 
by     $ r^{(k)}_{m,n; m',n'} $ with  any of  of the following four choices 
\eqn\choices{
m,n;m',n'=   \cases{m,n;1,1 & $ \qquad\qquad (1)$  \cr 
1,1;m,n &$ \qquad\qquad (2)$ \cr
1,n; m,1 &$ \qquad\qquad (3)$ \cr
m,1, 1,n &$ \qquad\qquad (4)$ }
}
 In the case of a non-perturbed theory ($t_k=0$),  
the first choice  was considered  in \refs{\KlebanovKM, \AlexandrovNN}
and the last two were
considered by Martinec \MartinecKA.

Our first-order calculation shows that this degeneracy is not lifted by 
perturbations by order operators. 
The matrix model does not distinguish between the four choices in
\choices\ and there is no convincing physical argument  that 
allows to single out one of them.

  It has been argued by the authors of \SeibergShih\ that in the
  $(p,q)$ string theory the $|m,n;1,1\rangle$ branes with $1\le m\le
  p-1$, $1\le n\le q-1$ and $mq-np >0$ form a complete set of distinct
  physical states with ZZ-type boundary conditions, which they called
  "principal branes".  The other branes should be thought of as
  multi-brane states formed out of these elementary ZZ branes.  The
  degeneracy we observed suggests that all four choices in \choices\
  provide a possible basis of "principal" branes and each of the
  choices describes the same set of physical states.
   
    This statement seems less strange if we recall the following two facts.
     First, it follows from the exact  expressions for  wave functions of 
     the FZZT and ZZ  boundary states 
    \refs{\FZZb, \ZamolodchikovAH},
     that  the ZZ state $\langle {m,n}|$ can be obtained as a difference of two FZZT states 
    $\langle \theta|$
\eqn\ZZFZZ{
\langle m,n|\sim \langle \th_{m,n}|-\langle \th_{m,-n}|
    =  \langle \th_{m,n}|-\langle \th_{-m,n}|  ,
}
where  the angle $\th$ is related to the boundary cosmological constant
    as $\mu_B\sim \sqrt{\mu} \cosh (p\theta)$ and 
    $\th_{m,n}$ is defined by \defsmn. 
    This 
    fact has been  further  explored in \refs{\Hos, \TeschnerQK, \Pons, \MartinecKA}.
    On the other hand, it is also known that
    the target space dimension in  the rational string theories is 
    associated with the imaginary direction of the uniformization 
    parameter $\th$.  Namely the points of the discrete target space 
    are labeled by the  $p-1$ cuts of the Riemann surface of the 
    function $Y(x)$ or the $q-1$ cuts of the Riemann surface of the 
    function $X(y)$. The   discontinuity  of $Y(x)$ along the $m$-th cut is
    given by $ -2\pi i 
    \rho^{(m)}(\th) = Y(\th + i\pi {m\over p} )- Y(\th - i\pi {m\over p} )$.
    From the point of view of CFT, the $m$-th cut describes a matter boundary condition of type $(m,1)$.
              Similarly, the 
              discontinuity  of $X(y)$ along the $n$-th cut is
    given by $ -2\pi i 
   \tilde  \rho^{(n)}(\th) = X(\th + i\pi {n\over p} )- X(\th - i\pi {n\over p} )$
   and describes the matter boundary condition $(1,n)$.
   The matter boundary conditions  $(m,n)$ has never been studied from the matrix point of view, but it  is plausible that they can be described in 
   terms of the  disc partition  function $\Phi(\th)$ 
   with  boundary parameter shifted  in the imaginary
  direction $\th\to \th +\th_{\pm m, \pm n}$. Thus it seems that  the  translations of the 
  boundary parameter by  $\th_{\pm m, \pm n}$ can be interpreted 
  either as projection to  $(m,n)$  ZZ boundary condition for the Liouville field, or as projection to  $(m,n)$  boundary condition for the matter field.
   Of course this statement needs  to be better understood.

\newsec{The $c\to 1$ limit and comparison with Matrix Quantum Mechanics} 

       \noindent
The $c=1$ limit is obtained by taking $b^2 \equiv {p\over p+1} \to 1$.
 In this limit  it is convenient to  use the  parameters 
$s$  and $M$ related to $\th$ and $\xi$ as   
\eqn\thsmxi{ \pi b s =  p\th , \quad M= \xi^p.}
The complex curve \CHEB\  
\eqn\curvepbig{ x(s) = M \cosh \pi b s, \quad y (s)= M^{1/b^2} \cosh \pi s/b}
becomes in the limit $p\to \infty$ 
\eqn\curvecone{ x(s)= M \cosh \pi s, 
\quad y (s) = x + {1\over p}[ x \log M+  \pi s M \sinh \pi s].}
 In this limit it is convenient  to redefine the  variable $y$ 
 by subtracting the  linear in $x$ term and rescaling by $1/p$:  
\eqn\newy{
   y (s) =   \pi s M \sinh \pi s .}
 The   function   \newy\  gives the continuum limit of the 
 resolvent in the   $\hat A\hat D\hat E$ matrix models \KostovIE, which  
 describe particular sectors of the $c=1$ string theory.

 In the description  based on the Matrix Quantum Mechanics \KlebanovMQM , 
 the natural variable is the canonical momentum $p$ 
 conjugated to the eigenvalue $x$, which is related to the resolvent by
 \eqn\pofx{p(x) = {y(e^{ i\pi} x) - y(e^{-i\pi } x)\over 2\pi i}}
 or, in terms of the parameter $s$,
    \eqn\MQMmom{ p(s) = { y( s+ i)-  y(s-i)\over 2\pi i}=M\sinh(\pi s)
    , \qquad 
    x = M\cosh(\pi s).}
 The   relation  
 \eqn\hypp{p^2 -x^2 = \mu, \qquad \mu = M^2,}
 is the equation for the   classical phase space trajectory 
 in MQM.   In the world-sheet CFT this relation appears in the context of the 
Witten's  ground ring \WittenZD.
 
    In the $c=1$ theory it is more natural to consider the chiral 
  variables
  \eqn\chiralv{ x_+= p+x,\quad x_-  = p-x}
  which describe the left and right moving tachyons.
  The  variables $x_\pm$ can be considered as a pair of local coordinates 
  covering the two-dimensional sphere.
  The punctures at the north and the south poles of the sphere 
  correspond to the points $x_+=\infty$ and $x_-=\infty$
       and the  two charts are related by the equation of the 
       complex curve \hypp
       \eqn\curvepm{
       x_+ x_- =\mu.}
As  in the case of the $(p,q)$ string theory, the perturbations  
are introduced as the asymptotics at the two punctures.
 The complex curve deformed
by  tachyon sources  $\sum_{n>0}t_{\pm n}  (x_\pm)^R$  is   found  in  
\AlexandrovNN. The analog of \dPhi\ is
  \eqn\dPhicone{\eqalign{
   d\Phi^+ &=  x_- dx_+ 
  +  \sum _{n>0} H_n  dt_n\cr
d\Phi^- &=  x_+ dx_- 
 + \sum _{n<0} H_n dt_n ,}
 }
 where the Hamiltonian $H_{\pm n}$ 
 is  given by the positive/negative part of the 
 Laurent expansion of the function $x_\pm^n(\o)$, $\o = e^{\pi s}$,
 plus half the constant term. 
 Note that here we should keep  both  expansions 
 since  the singular points $x_+=\infty$ and $x_-=\infty$  are distinct.
  The  parameter $\o$ is expressed as a function of 
  $x_+$ or $x_-$ as
  $ \o  = e^{i \p_\mu \Phi_+(x_+)}=e^{i \p_\mu \Phi_-(x_-)}
  .$
 The  conditions  \FPERTs\ and \NORMAs\ 
  turn to
  \eqn\FPERMQM{  \eqalign{
  \p_\mu\CF&= \int_{B} x_-dx _+= \  \oint_B d\Phi ^+,
  \cr  \mu&= \int_{A} x_-dx_+= \quad \oint_A d\Phi ^+ ,} 
  }
   where the compact cycle $A$ goes around the equator and the 
   non-compact cycle 
   $B$ connects the 
   two    punctures \AKKnmm.

   Now let us concentrate on the instanton corrections in MQM
   compactified at radius $R$ and their CFT interpretation.  In the
   non-perturbed theory described by \curvepm\ the non-perturbative
   corrections follow from the integral representation of the free
   energy \KlebanovMQM \eqn\nonpernp{ \CF= \CF^{\rm pert} +
   \sum_{n=1}^\infty C_n e^{ -2\pi n \mu} +\sum_{n=1}^\infty \tilde
   C_n e^{ - 2\pi n R\mu }.}  The world-sheet description of the two
   kinds of exponential terms was discussed in \AlexandrovNN. The
   terms $ e^{-2\pi n \mu}$ correspond to Dirichlet boundary
   conditions for the matter field.  Due to the translational symmetry
   of the target space, the instantons are labeled by only one number
   $n$ (the other one is redundant).  The integral \Smn\ is replaced
   by
 an integral along  a cycle 
going $n$ times around the neck of the hyperboloid.  Thus the 
$n$-instanton solution is associated with a contour winding $n$ times 
around the neck.  The original 
algebraic curve  is actually the universal 
covering of the hyperboloid, due to the logarithmic dependence of 
$\hat y (x)$.
  The curve wraps   infinitely 
many times the hyperboloid $x^+x^-=\mu$. 
Similarly,  the  
terms $e^{ - 2\pi n\mu R}$ correspond to Neumann
boundary conditions for the matter field
and correspond to the cycles of the dual curve describing the vortex excitations. 
 It was argued in   \AlexandrovUN\
that the Liouville boundary conditions for the $n$-th instanton correction
are given by the $(1,n)$ ZZ boundary state. The argument of 
 \AlexandrovUN\ was based  on  the comparison of the  
 first order perturbations in MQM and the world sheet CFT.

 Here we will show that the instanton corrections found in 
  \refs{\AlexandrovNN,  \AlexandrovUN} are related to the 
  disc partition function by a formula similar to \Ssmn.
  This statement is trivial when the theory is not perturbed.
  Indeed,  in this case
  \eqn\xmncone{\eqalign{
  x_\pm &= \mu^{1/2} \o^{\pm 1} = \mu^{1/2} e^{\pm \pi s}
  ,\cr
   \Phi^+(x_+)&= \mu \log (\mu^{-1/2}x_+) = \mu \pi s - \hf \mu \log\mu
  }
  }
  and 
  \eqn\Phimncone{ \Phi_+(s=in  )-\Phi_+(s=-i\pi n) =  2\pi i n \mu,}
    which reproduces the first kind of instanton corrections.
    By construction the potential $\Phi_\pm (x_\pm)$ are the
    generating functions of the tachyon operators and 
    as such describe Dirichlet  branes with fixed time position.
       The second kind    of instanton corrections
       is reproduced by the effective potential $\tilde \Phi_\pm$ in the 
       dual theory, which describes vortex  excitations and therefore 
       Neumann  branes wrapping the  time circle.

       Below we  will consider  the effect of the  tachyon perturbations 
       with  $\tilde t_1=\tilde t_{-1}=\lambda, \ t_{\rm others}= 0$, 
       which affect only the Dirichlet branes.
       Then the instanton corrections take the form \refs{\AlexandrovNN,  \AlexandrovUN} 
          $$ \Delta \CF=\sum_{n=1}^\infty C_n  e^{ - \mu  f_n(\mu,\lambda)} 
  +\sum_{n=1}^\infty \tilde C_n  e^{ -2\pi nR \mu
   }$$
   with
  \eqn\fenn{   f_n(\mu,\lambda)= 
2\pi n    +4 \lambda\, {{\sin(\pi n/R) } }\, \mu^{{1\over 2R}-1}\, +
{\lambda^2\over R^2} \, {\sin(2\pi  n/R) } \, \mu^{{1\over R} -2}\,+
 ...
}
The   exponent \fenn\   
can be presented again as 
\eqn\fenPh{
\mu f_n (\mu,\lambda)={\Phi_+( s= i n)-\Phi_+(s=-i n) \over i}}
  where
\eqn\PHIP{
\Phi_+(x_+)
=   \mu\log (\mu^{-1/2}x_+) -2\lambda\, \mu^{{1\over R}}\, (x^+)^{-{1\over R}}
-   {   \lambda ^2 \over 2R^2} \,\mu^{{2\over R}-1}\,  (x^+)^{-{2\over R}}+ ...
}
is the effective potential in presence of perturbation.
  It can be formally  interpreted as  the disc partition function with 
  a ``chiral"  (and therefore non-local) 
  boundary condition  labeled by the value of $x_+$.
      The  relation \fenPh\ can be generalized to the case of  finite 
      perturbation \insts.

      \newsec{Discussion} 
  
In this work we  derived the
non-perturbative corrections in the unitary $(p,p+1)$
models of 2D quantum gravity  from the collective  field  theory 
for the two-matrix model.  We used the  approach developed   in
\refs{\DavidSK,\DavidZA}  for the one-matrix model.  The results 
have a nice algebro-geometric interpretation: the contribution of an
instanton corresponds to the integrals over the cycles starting and
ending at two different sheets of the Riemann surface of the algebraic
curve of the model, at the points corresponding to a pinched cycle
connecting two sheets.  In
the CFT description this interpretation was proposed in \SeibergShih\
where the curve appeared from the ground ring relations.

We also compared the new results with the Liouville CFT (generalizing
the successful comparison done in
\AlexandrovNN\ (see also \MartinecKA\ where the idea of such a comparison 
was proposed).  The CFT interpretation of our matrix instanton
calculation is the following: we computed the one point functions of
primary fields on the ZZ brane already found on the CFT side in
\ZamolodchikovAH. As usual, the normalizations of operators are 
different in matrix and CFT approaches, and it only makes sense to
compare the dimensionless ratios of various quantities. For that we
had to calculate the free energy and the two point functions of the
primary fields on the sphere in the same instanton approach. This
calculation appears to be subtle due to the singular nature of the
critical algebraic curve. We succeeded to do it using the geometry
relations \refs{\DavidSK,\DavidZA, \DijkgraafFC}  defining the free energy of
the model through holomorphic integrals over the curve, combined with
the bilinear Riemann identity. 

Our method may be not as straightforward as the old method based on
the string (KP) equations but it is more transparent geometrically and
might have a wider range of applications, especially for the matrix
models where the method of orthogonal polynomials does not exist but
the curve is known.

The instanton method also clarifies the geometrical meaning of the
relation between FZZT brane and ZZ brains, as mentioned in
\refs{\SeibergShih, \TeschnerQK}: they correspond to different choices of the 
contour in the same holomorphic integral: for the FZZT brane the
contour starts at any point on the curve (corresponding to the complex
boundary cosmological constant) and goes to infinity, whether as for
the ZZ brane the contour connects two different critical 
points  corresponding to the pinched cycle of the  algebraic
curve.

Our  results show that the agreement between the  matrix model and CFT calculations  established in \AlexandrovNN\  for the $(p,p+1)$ critical points, 
holds  also  in presence of  perturbations  by order operators,
at least in the linear order in the couplings.  
  The most interesting physical outcome of our  calculation is that the
results do not depend on the choice of the ZZ branes: $(m,n)$,
$(m,1)$, $(1,n)$ and $(1,1)$ ZZ branes give the same instanton effects
being combined with $(1,1)$, $(1,n)$, $(m,1)$ and $(m,n)$ matter
branes, respectively.  This degeneracy  is needed for the self-consistency of  the matrix interpretation
of the ZZ branes as non-perturbative effects since the difference in the
results would mean that we were missing the matrix model description
of some of these branes. This degeneracy is still to be  understood
within the Liouville CFT of 2D gravity.

We also applied our method to the study of the instanton effects the
$c=1$ string compactified on an arbitrary radius and perturbed by
relevant vortex operator. We showed that the instantons have here
essentially the same geometrical meaning in terms of the holomorphic
integrals along the algebraic curve as for $c<1$ models. We also
identified the multi-instanton contributions in this case as multiple
windings of the integration contour along the cycles of the curve.
It is known that this model gives the black hole realization proposed
in \KazakovPM.  The non-perturbative corrections play there the
crucial role for the understanding of black hole physics in the near
horizon strongly coupled area. Our method opens the way for the
geometrical instanton interpretation of the energy of the ZZ-brane in
the black hole calculated in \AlexandrovNN\ by the matrix approach
using Toda equations of \KazakovPM.

An interesting continuation of our approach would be the
generalization of the results to the models perturbed far away from
the simple critical curve \CHEB. As we showed in section 2, the
instanton approach works well for any two matrix model, for a general
potential.  Thus our methods are directly applicable to study the
non-perturbative effects in the flows between different minimal
models.

In conclusion, the non-perturbative effects in the large $N$ matrix
models seem to find their most natural and universal interpretation in
the eigenvalue instanton method studied here. Its application range
should be much wider then the methods based on orthogonal polynomials
and could be useful everywhere where the model can be described in the
quasi-classical limit by its algebraic curve.

\bigskip 
\noindent{\bf Acknowledgments:} 
We thank S.~Alexandrov, F. ~David, D.~Kutasov and  A.~Marshakov  for useful
discussions and to N.~Seiberg, D.~Shih and P.~Wiegmann, for their
critical remarks on the manuscript.  We would like also to thank the
Max Planck Inst. (Potsdam) for the kind hospitality during the course
of this work.  The work of V.K. was partially supported by European
Union under the RTN contracts HPRN-CT-2000-00122 and -00131, by NATO
grant PST.CLG.978817 and by the ECOS-Sud-C01E05 grant, during the stay
of V.K. at the Catolica Univ. (Chile). The work of I.K. was also
supported in part by European networks European networks EUROGRID
HPRN-CT-1999-00161 and  EUCLID HPRN-CT-2002-00325.

\bigskip 
 \centerline{***}
\bigskip

\noindent 
{\sl We started this work when our friend Ian Kogan was still
alive. We remember well our happy gatherings at this time in his
apartment in Bures-sur-Yvette during his stay in IHES, discussing
physics, but also singing, laughing, drinking together. He attracted
many different people by his generous friendliness.  We hope 
 that a spark of his generosity and friendship is
reflected in our present work.  We will always remember him.}


 \listrefs
 \bye